\documentclass[fleqn,usenatbib]{mnras}

\usepackage{newtxtext,newtxmath}

\usepackage[T1]{fontenc}

\DeclareRobustCommand{\VAN}[3]{#2}
\let\VANthebibliography\thebibliography
\def\thebibliography{\DeclareRobustCommand{\VAN}[3]{##3}\VANthebibliography}

\usepackage{graphicx}	
\usepackage{amsmath}	
\usepackage{amssymb}	
\usepackage{xcolor}
\usepackage{comment}
\usepackage{soul}
\usepackage{float}

\newcommand{\Msun}{M_\odot}
\newcommand{\Mdot}{\dot{M}}
\newcommand{\Mdotstar}{\dot{M}_\star}
\newcommand{\Mdotin}{\dot{M}_\mathrm{in}}
\newcommand{\Mdotcrit}{\dot{M}_\mathrm{crit}}

\newcommand{\Pdot}{\dot{P}}

\newcommand{\rin}{r_\mathrm{in}}

\newcommand{\rco}{r_\mathrm{co}}

\newcommand{\rinmax}{r_\mathrm{in,max}}
\newcommand{\rA}{r_\mathrm{A}}
\newcommand{\reta}{r_\eta}
\newcommand{\rxi}{r_\xi}

\newcommand{\Rin}{R_\mathrm{in}}
\newcommand{\RA}{R_\mathrm{A}}
\newcommand{\Reta}{R_\eta}
\newcommand{\Rxi}{R_\xi}
\newcommand{\Rinmax}{R_\mathrm{in,max}}
\newcommand{\DeltaR}{\Delta r/\rin}

\newcommand{\Ostar}{\Omega_\star}

\newcommand{\Lx}{L_\mathrm{X}}

\newcommand{\Porb}{P_\mathrm{orb}}
\newcommand{\Porbi}{P_\mathrm{orb,i}}

\newcommand{\Pbif}{P_\mathrm{bif}}
\newcommand{\Prlof}{P_\mathrm{RLOF}}

\newcommand{\gamamb}{\gamma_{\rm{MB}}}
\newcommand{\mesa}{{\tt MESA}}

\newcommand{\gpers}{g~s$^{-1}$}  
\newcommand{\ergpers}{erg~s$^{-1}$}

\newcommand{\spers}{s~s$^{-1}$}

\newcommand{\x}{\times}

\title[Binary and NS evolution in LMXBs along AMXP tracks]{Binary and neutron star evolution in low-mass X-ray binaries on the evolutionary tracks of accreting millisecond X-ray pulsars}

\author[Niang et al.]{
N. Niang,$^{1}$\thanks{E-mail: niang@sabanciuniv.edu}
\"{U}. Ertan,$^{1}$
A. A. Gen\c{c}ali,$^{1}$
F. Ertuğrul,$^{1}$
A. Ulubay$^{2}$,
E. Devlen$^{3}$
and 
M. A. Alpar$^{1}$
\\
$^{1}$Sabanc{\i} University, Orhanl{\i}, Tuzla, 34956, \.{I}stanbul, Turkey\\
$^{2}$Faculty of Science, Department of Physics, Istanbul University, 34134, Vezneciler, \.{I}stanbul, Turkey\\
$^{3}$Faculty of Science, Department of Astronomy and Space Sciences, Ege University, 35100, Bornova, \.{I}zmir, Turkey\\
}

\date{Accepted XXX. Received YYY; in original form ZZZ}

\pubyear{\the\year{}}

\begin{document}
\label{firstpage}
\pagerange{\pageref{firstpage}--\pageref{lastpage}}
\maketitle

\begin{abstract}
Neutron star low-mass X-ray binaries (LMXBs) are the progenitors of millisecond pulsars. In these systems, old neutron stars (NSs) can be spun up during a long-lasting accretion phase. The discovery of accreting millisecond X-ray pulsars (AMXPs) and transitional millisecond pulsars has provided key observational insights into the connection between millisecond pulsars and LMXBs. In this work, we have investigated both the binary system and the individual NS evolution leading to AMXP properties. We use \mesa~to analyse the binary evolution of LMXBs, following three distinct evolutionary tracks defined by the AMXP donor types. We find that while the magnetic braking index may affect the mass-transfer history, the initial orbital period is the most influential parameter that shapes the overall binary evolution. We use the mass accretion histories estimated from these binary simulations to study the rotational evolution of NSs employing the model that can account for torque-luminosity relations and the lack of X-ray pulses from most of these systems. With reasonable model parameters, our model results are in agreement with the typical properties of AMXPs. For these AMXP sources from each evolutionary track, we have shown that the model can reproduce the NS and binary properties simultaneously. Finally, we discuss the time-scales of different evolutionary paths, as well as the conditions under which these systems could be detectable at various stages of their evolution.

\end{abstract}

\begin{keywords}
accretion, accretion discs, stars: neutron
\end{keywords}



\section{Introduction}
In neutron star low-mass X-ray binaries (LMXBs), the compact object accretes matter from its low-mass companion with mass $M_2 < 1 ~ M_{\odot}$ through Roche-lobe overflow (RLOF) in a close binary (with orbital period $\Porb < 1~$d). NSs in LMXBs can be spun up to millisecond periods through accretion torques \citep{Alpar1982,Radhakrishnan1982ONPULSAR}. This process, so-called \textit{recycling scenario}, positions the LMXBs as the progenitors of weakly magnetized radio millisecond pulsars (RMSPs)  \citep[for a review see][]{Srinivasan2010}. First observational support for this scenario came with the discovery of coherent $2.5~$ms X-ray pulsations from the first accreting millisecond X-ray pulsar (AMXP) SAX 1808.4-3656 \citep{Wijnands1998ASystem}. At present, there are $23$ known AMXPs \citep{Patruno2021AccretingPulsars}. The recently discovered three transitional millisecond pulsars (tMSPs), which show transitions between the X-ray pulsar and radio pulsar states, have provided the missing link connecting LMXBs to RMSPs \citep{Archibald2009ALink,Papitto2013SwingsPulsar,Bassa2014AXSSJ122704859, Papitto2022TransitionalPulsars}. 

The long-term evolution of LMXBs can be summarized as follows: (i) In the first stage, the source is spun up by the mass accretion on to the NS (at rates close to the Eddington limit). The accretion also causes the rapid decay of the magnetic field through ohmic dissipation \citep{geppert1994,urpin1995,konar1997,srinivasan1990}. The dipole field is expected to freeze at a strength of a few $10^{7} - 10^{9}~$G in correlation with the accretion rate in the early stages of the evolution \citep{konar1997, Srinivasan2010}. (ii) In the subsequent phase, the mass inflow rate of the disc, $\Mdotin$, decreases and the system is likely to be observed as a transient source like AMXPs. With further decrease in $\Mdotin$, the binary goes through a Roche-lobe decoupling phase \citep{Tauris2012}. (iii) In the final LMXB phase with the lowest $\Mdotin$ levels, the system could become a tMSP, and eventually becomes an RMSP after the NS enters the strong propeller phase.

The observed properties of AMXPs summarized in Table \ref{TableAMXP} categorize them into three distinct groups: The first group consists of $9$ sources with main sequence (MS) companions of mass $0.1~\Msun < M_2  < 0.5~\Msun$ and $3~$h $< \Porb < 19~$h. These sources are the fastest spinning AMXPs with median spin period $ \langle P \rangle ~\simeq ~2.2~$ ms. The second group includes $4$ sources with low-mass ($M_2 \simeq 0.01 \Msun$) degenerate brown dwarf (BD) companion and $1~$h $< \Porb < 3~$h with $ \langle P \rangle ~\simeq ~2.5~$ ms. The last group includes AMXPs in ultra-compact X-ray binaries (UCXBs) with $\Porb < 80~$min. These are $10$ sources with extremely low-mass ($M_2 \simeq 0.001 \Msun$) degenerate He/CO white dwarf (WD) companions. These systems with $ \langle P \rangle ~\simeq ~5.2~$ ms, spin relatively slowly compared to other AMXPs. The categorization of AMXPs into these three groups suggests distinct evolutionary paths for their LMXB progenitors. This has been suggested by \cite{Deloye2008ThePerspective}, who explored the evolutionary connections between LMXBs and RMSPs in close orbits. A critical aspect influencing these formation channels is the bifurcation period $\Pbif$. This is the value of $\Porb$ at the onset of RLOF and is estimated to be $\Pbif \lesssim 1~$d \citep{Pylyser1988EvolutionComponent.,Pylyser1989TheLosses.}. LMXBs with $\Porb < \Pbif$ are converging systems, which evolve with decreasing $\Porb$. In contrast, the binaries with $\Porb > \Pbif$ are diverging systems evolving into wide-orbit systems with $\Porb > 10~$d. 

 The evolutionary tracks for the AMXP groups with BD, He WD, and MS companions described above can be referred to as BD-track, UCXB-track, and MS-track. The details of these evolutionary tracks were investigated by \cite{He2019FormationPulsars}, who successfully reproduced most of the binary and companion properties of AMXPs ($\Porb, M_2$, and the X-ray luminosity $L_{X}$). Other studies have also investigated the evolution of binary and companion properties, as well as the formation of RMSPs, UCXBs, and other AMXPs  \citep{Istrate2014ThePeriod, Sengar2017NovelStars,Tauris2018DisentanglingLISA}. Formation of AMXPs considering the evolution of both the binary and the individual properties of the two stars was first studied by \cite{Kar2024Long-termPulsars}. There are other recent works that studied the long-term evolution of LMXBs by integrating the spin evolution of the NS with a focus on the formation of RMSPs and spider pulsars, skipping the AMXP stage \citep{Lan2023ThePulsars, Misra2025TheTransfer, Misra2025InvestigatingEvolution}. In this work, we will use the stellar evolution code Modules for Experiments in Stellar Astrophysics \citep[{\tt MESA},][]{Paxton2011ModulesMESA,Paxton2013ModulesStars,Paxton2015ModulesExplosions,Paxton2018ModulesExplosions,Paxton2019ModulesConservation, Jermyn2023ModulesInfrastructure} to investigate the long-term evolution of LMXBs with initial orbital periods below or near $\Pbif$. These systems undergo long-lasting, stable mass transfer, leading to the formation of AMXPs following three specific evolutionary tracks: BD-track, UCXB-track, and MS-track. For each of these evolutionary tracks, we will also study the rotational evolution of the NS using the analytical model developed by \cite{Ertan2021}.

The duration and stability of mass transfer from the companion to the Roche lobe (RL) of the NS depend significantly on the mechanisms of mass and angular momentum loss (AML). In LMXBs, AML is primarily driven by the gravitational radiation (GR) and magnetic braking (MB) \citep{Bhattacharya1991FormationPulsars,Tauris2023PhysicsSources}. The AML by GR is efficient particularly  in tight-orbit systems like UCXBs during the late phases of evolution. Conversely, MB can efficiently drain the orbital angular momentum of the binary during the early phases of the evolution. MB occurs in Sun-like MS or dwarf stars with convective envelopes. As the donor star slows down, the angular momentum is carried away by a magnetically coupled stellar wind, with a negligible mass loss of the donor star. Recent studies have shown that the conventional MB model, commonly referred to as the Skumanich law \citep{Skumanich1972TimeDepletion,Rappaport1983ABraking.}, is not successful in reproducing some observed properties of LMXBs \citep{Pavlovskii2016MassX-1}. In particular, the mass transfer rates are underestimated by an order of magnitude with these conventional models  \citep[see][for details]{Van2019Low-massPrescription}. Formation and evolution of accreting LMXBs were also investigated with different MB models \citep{Deng2021EvolutionPrescriptions,Echeveste2024TheLMXBs,Yang2024TheBinaries}. \cite{Deng2021EvolutionPrescriptions} analysed the effect of  MB prescription on both LMXB and RMSP stages, combining binary evolution and binary population synthesis. The results of this work indicate that the convection-boosted ($\tau$- boosted) MB model proposed by \cite{Van2019Low-massPrescription} is a suitable MB model to study the long-term evolution of LMXBs. Based on these findings, we adopt the $\tau$-boosted MB model to investigate the evolutionary tracks of AMXPs with {\tt MESA}.

The analytical model employed in this work is based on the basic principles of the disc–field interaction model developed by \cite{Lovelace1999MagneticOutflows} and \cite{Ustyugova2006PropellerStars}. In this model, the closed field lines that are interacting with the inner disc within a narrow boundary layer inflate, open up, and reconnect on the dynamical timescale of the disc. The field lines are open and disconnected from the outer disc outside the interaction region. In the conventional models, the entire disc is assumed to be threaded by the closed field lines inside the light cylinder \citep{Ghosh1979ACCRETION1}. However, it was shown in later work that the closed field lines rotating with the star cannot slip through the disc since the diffusion timescale of the field lines is much longer than the interaction timescale of the lines with the inner disc \citep{Fromang2009TurbulentInstability}. At the inner disc radius, ${\rin}$, the field lines should be sufficiently strong to force the matter into co-rotation within the short interaction timescale. In the strong propeller (SP) phase, all the inflowing matter can be thrown out of the system along the open field lines \citep[see e.g.,][]{Uzdensky2004MagneticDisks}. There is a maximum radius at which this SP mechanism can work, which is likely to be close to the actual $\rin$. \citet{Ertan2017TheStars} estimated this $\rin$ through analytical calculations as a function of $P$, $\dot{M}_{\rm in}$, and the magnetic dipole moment $\mu$, and found that it is much smaller than the conventional Alfvén radius, $\rA$, in this phase. The model was later extended to calculate $\rin$ and the torques in all rotational phases (SP, weak propeller (WP), and spin-up (SU)), as well as the critical conditions for the transitions between these phases \citep{Ertan2021}.

Conventional models, which assume the inner disc radius to be close to $\rA$, cannot account for some typical properties of LMXBs. These include ongoing accretion at low $\Lx$ levels while the NS spins down, as well as torque reversals occurring with small changes in $\Lx$ and with comparable torque magnitudes on either side of the reversal \citep[for a review see][]{Bildsten1997OBSERVATIONSPULSARS}. Applications of the model of \citet{Ertan2017TheStars, Ertan2021} to different classes of LMXBs yielded results that are consistent with observations. As a strongly magnetized system, the torque reversals of 4U 1626–67 were studied by \citet{Gencali2022}. In weakly magnetized systems, the model was applied to investigate the SP / WP transitions of tMSPs \citep{Ertan2018} and to explain the lack of X-ray pulsations from most LMXBs, while a small fraction exhibit X-ray pulsations as AMXPs \citep{Niang2024OnBinaries}.

     During the long-term evolution, the torques acting on the NS depend sensitively on both $\dot{M}_{\rm in}$ and $\rin$. \citet{Kar2024Long-termPulsars} provided a comprehensive study of the long-term evolution of AMXPs. They reported that reproducing the observed spin frequencies of the fast-spinning AMXPs ($P \lesssim 3~$ms) with degenerate WD/BD companions remains challenging, and suggested that this problem could be resolved by considering the effects of transient accretion. We also use \mesa~for the binary evolution and obtain $\Mdotin$ histories similar to those in \citet{Kar2024Long-termPulsars}. The fact that our simulations can reproduce the properties of these fast rotating AMXPs indicates that the basic reason leading to different fast-spinning NS properties between the two analyses is the difference in the employed torque models together with $\rin$ calculations. We adopt the analytical model developed by \citet{Ertan2021} whereas \citet{Kar2024Long-termPulsars} used the torque model of \citet{Bhattacharyya2017THEPULSARS}.

 In this work, our aim is to investigate the binary evolution of LMXBs together with the rotational evolution of their NSs until the sources reach their equilibrium spin periods. For these calculations, we will use {\tt MESA} to study the details of the evolutionary paths of AMXPs (BD-track, UCXB-track, and MS-track) and obtain the evolution of their  donor's long-term mass transfer rate $\Mdot_2$. Using the $\Mdot_2$ histories estimated by \mesa~simulations, we will study the rotational evolution of the NS using the aforementioned analytical model. The evolution of the binary and donor properties ($\Porb, M_2$, and $\Mdot_{2}$) for each track is described in Section \ref{Section2}. We briefly describe the analytical model together with the illustrative long-term rotational evolutions of the NS (evolution of $P$ and $\Pdot$) evolving along each track in Section \ref{Section3}. We apply our results to selected individual AMXPs from the three evolutionary tracks in Section \ref{Section4}. These sources, namely HETE J1900.1-2455 for the BD-track, XTE J1751-305 for the UCXB-track, and Aql X-1 for the MS-track, have been observed over a long period and/or throughout multiple outbursts. We discuss our results in Section \ref{Section5} and summarize our conclusions in Section \ref{Section6}.

\begin{table*}
\centering
\caption{ Observed properties of 23 AMXPs. The distance, $d$, and  $L_\mathrm{X,avg}$ values are obtained from a catalogue of outbursting NSs $(1)$. The $L_\mathrm{X,avg}$ values correspond to the average of $L_\mathrm{X,peak}$ values reached during each outburst, and $\dot{M}_{\rm avg} = L_{\rm X, avg} r_{\star}/{GM_1}$. The companion masses $M_{\rm 2,min}$ are taken from the recently released catalogue of LMXBs  $(2)$. In the seventh column we note some of the features of the sources (MS: main sequence, BD: brown dwarf, U: ultra-compact X-ray binaries, B: X-ray bursters, PRE: photospheric radius expansion observed during X-ray burst, GC: sources in globular clusters).}
\label{TableAMXP}
\resizebox{0.99\textwidth}{!}{%
\begin{tabular}{lccccccll}
\hline\hline
{\bf Name} &  $P_\mathrm{orb}$ &  {$ P$} &  d &  $L_\mathrm{X,avg}$ & $\dot{M}_\mathrm{avg}$ &  {$M_{\rm 2,min}$} & {Flags} & {References}\\
 & {(h)} & {(ms)} & (kpc) & ( \ergpers ) & ( \gpers ) & {\bf ($M_{\odot}$)} & & \\
\hline
Aql X-1                     & 18.95  & 1.82  & 4.50  & $1.94 \times 10^{37}$  & $1.04\times 10^{17}$  & 0.56     & MS,B         & $(1)$, $(2)$ \\
SWIFT J1749.4-2807          & 8.82   & 1.93  & 6.70  & $1.67 \times 10^{36}$  & $8.97\times 10^{15}$  & 0.59     & MS,B         & $(1)$, $(2)$ \\
IGR J17591-2342             & 8.80   & 1.90  & 8.00  & $2.50 \times 10^{36}$  & $1.35\times 10^{16}$  & 0.42     & MS           & $(1)$, $(2)$ \\
SAX J1748.9-2021            & 8.76   & 2.26  & 8.50  & $2.60 \times 10^{37}$  & $1.40\times 10^{17}$  & 0.10     & MS,B,GC      & $(1)$, $(2)$ \\
SRGA J144459.2-604207       & 5.22   & 2.23  & 8.50  & $1.32 \times 10^{37}$  & $7.10\times 10^{16}$  & 0.2 - 0.7 & MS,B        & $(1)$, $(2)$ \\
MAXI J1816-195              & 4.83   & 1.89  & <6.3  & $1.50 \times 10^{37}$  & $8.07\times 10^{16}$  & 0.55     & MS,B         & $(1)$, $(2)$ \\
XTE J1814-338               & 4.27   & 3.20  & 8.00  & $1.59 \times 10^{36}$  & $8.56\times 10^{15}$  & 0.17     & MS,B,PRE     & $(1)$, $(2)$ \\
IGR J17498-2921             & 3.84   & 2.49  & 7.60  & $3.85 \times 10^{36}$  & $2.07\times 10^{16}$  & 0.17     & MS,B,PRE     & $(1)$, $(2)$ \\
IGR J17511-3057             & 3.47   & 4.08  & 6.90  & $2.70 \times 10^{36}$  & $1.45\times 10^{16}$  & 0.13     & MS,B         & $(1)$, $(2)$ \\
\hline    
IGR J00291+5934             & 2.46   & 1.67  & 4.20  & $1.12 \times 10^{36}$  & $6.00\times 10^{15}$  & 0.039    & BD           & $(1)$, $(2)$ \\
SAX J1808.4-3658            & 2.01   & 2.49  & 3.50  & $1.69 \times 10^{36}$  & $9.07\times 10^{15}$  & 0.043    & BD,B,PRE     & $(1)$, $(2)$ \\
IGR J17379-3747             & 1.88   & 2.14  & 8.50  & $2.48 \times 10^{36}$  & $1.34\times 10^{16}$  & 0.056    & BD           & $(1)$, $(2)$ \\
HETE J1900.1-2455           & 1.39   & 2.65  & 4.70  & $2.10 \times 10^{36}$  & $1.13\times 10^{16}$  & 0.016    & BD,B         & $(1)$, $(2)$ \\
\hline
IGR J17494-3030             & 1.25   & 2.66  & 8.00  & $8.85 \times 10^{35}$  & $4.76\times 10^{15}$  & 0.02     & He WD?, U    & $(1)$, $(2)$ \\
MAXI J1957+032              & 1.01   & 3.19  & 5.00  & $1.43 \times 10^{36}$  & $7.69\times 10^{15}$  & 0.017    & He WD?       & $(1)$, $(2)$, $(3)$ \\
NGC 6440 X-2                & 0.96   & 4.86  & 8.50  & $1.43 \times 10^{36}$  & $7.69\times 10^{15}$  & 0.00067  & He WD, U, GC & $(1)$, $(2)$  \\
SWIFT J1756.9-2508          & 0.91   & 5.49  & 8.00  & $2.78 \times 10^{36}$  & $1.49\times 10^{16}$  & 0.007    & He WD, U     & $(1)$, $(2)$ \\
IGR J16597-3704             & 0.77   & 9.51  & 9.10  & $2.20 \times 10^{36}$  & $1.18\times 10^{16}$  & 0.006    & He WD, U,GC  & $(1)$, $(2)$ \\
MAXI J0911-655              & 0.74   & 2.94  & 9.50  & $1.90 \times 10^{36}$  & $1.02\times 10^{16}$  & 0.024    & He WD?, U,GC & $(1)$, $(2)$ \\
XTE J0929-314               & 0.73   & 5.40  & >6    & $6.24 \times 10^{36}$  & $3.36\times 10^{16}$  & 0.0083   & C/O WD, U    & $(1)$, $(2)$ \\
XTE J1751-305               & 0.71   & 2.30  & 8.50  & $6.24 \times 10^{36}$  & $3.36\times 10^{16}$  & 0.014    & He WD, U     & $(1)$, $(2)$ \\
XTE J1807-294               & 0.67   & 5.25  & 8.00  & $9.60 \times 10^{36}$  & $5.17\times 10^{16}$  & 0.0066   & C/O WD, U    & $(1)$, $(2)$ \\
IGR J17062-6143             & 0.63   & 6.11  & 7.30  & $8.70 \times 10^{35}$  & $4.68\times 10^{15}$  & 0.0006   & He WD?, U,B  & $(1)$, $(2)$ \\
\hline
\end{tabular}
}\\
{Note. $(1)$ : \citealt{Heinke2025CatalogBinaries}, $(2)$: \citealt{Avakyan2023XRBcats:Catalogue}, $(3)$ \citealt{Sanna2022MAXIJ1957+032:Binary}}

\end{table*}

\section{PHYSICAL
ASSUMPTIONS AND NUMERICAL setup for the binary evolution }
\label{Section2}

\subsection{Angular momentum loss mechanisms}
\label{sec:AML}

We calculate the rate of orbital angular momentum loss ${\dot{J}}_{\rm{orb}}$ as

\begin{equation}
    \frac {{\dot J}_{\rm{orb}}}{{J}_{\rm{orb}}} = \frac {{\dot J}_{\rm ML}}{{J}_{\rm{orb}}}+\frac {{\dot J}_{\rm GR}}{{J}_{\rm{orb}}}+ \frac {{\dot J}_{\rm MB}}{{J}_{\rm{orb}}}
    \label{eq:Jdot_orb}
\end{equation}

where the terms on the right-hand side of this equation represent the orbital AML mechanisms by means of mass loss (ML), GR, and MB, respectively. 

\subsubsection*{Mass loss from the binary system}
\label{subsec:ML}

The orbital AML due to ML is commonly given as

\begin{equation}
    \frac{\dot{J}_{\mathrm{ML}}}{J_{\text {orb }}}=\frac{\alpha + \beta q^{2} + \delta\gamma (1+q^2)}{1+q} \frac{\dot{M}_{2}}{M_{2}}~{\text s}^{-1}
    \label{eq:ML1}
\end{equation}

\citep{Tauris2023PhysicsSources}, where $\alpha$ is the coefficient for ML through wind from the companion, $\beta$\footnote{ We note that this definition of $\beta$ differs from the one used among binary modelling works (e.g. \citep{Podsiadlowski2002EvolutionaryBinaries})} determines the fraction of matter ejected from the vicinity of the NS \citep[isotropic re-emission model,][]{Bhattacharya1991FormationPulsars}, $\delta$ is the coefficient for ML from a circumbinary co-planar disc with radius $a_{\rm r} = \gamma^2a$, where $a$ is the orbital separation. $q = M_2 / M_1$, $M_1$ is the NS mass, and $\Mdot_2$ is the mass-transfer rate of the donor. The ML through the wind is negligible and there is no circumbinary disc around these systems; therefore we take $\alpha = \delta = 0$. Equation (\ref{eq:ML1}) then reduces to:

\begin{equation}
    \frac {\dot{J}_{\mathrm{ML}}}{{J}_{\rm{orb}}} = \beta~\frac {q^2}{1+q}~\frac {{\dot M}_2}{M_2}~{\text s}^{-1}
        \label{eq:ML2}
\end{equation}

 The accretion efficiency given by $\epsilon = 1-\alpha- \beta -\delta$ reduces to $\epsilon = 1 - \beta$. As a result, the mass-transfer rate onto the NS is $\Mdot_1 = \epsilon~ |\Mdot_2|$.

\subsubsection*{Gravitational radiation}
\label{subsec:GR}

The orbital AML due to GR can be written as 

\begin{equation}
    \frac {{\dot J}_{\rm GR}}{{J}_{\rm{orb}}}= -\frac {32}{5~c^5}~\frac {G^{3}{M_1}~{M_2}~({M_1}+{M_2})}{a^4}~{\text s}^{-1}
        \label{eq:GR1}
\end{equation}

\citep{Faulkner1971Ultrashort-PeriodCamelopardalis}, where  $G$ is the gravitational constant, and $c$ is the speed of light. In close-orbit systems ($\Porb < 6~$h), which have degenerate companion stars, GR is the dominant AML mechanism due to the $a^{-4}$ dependence in equation (\ref{eq:GR1}). LMXBs in which GR is the dominant torque correspond to systems with $\Porb < \Pbif$ that are in an advanced phase of evolution. As GR is not effective during the early stages of the evolution, there must be another torque that drains the angular momentum of the system, which is linked to the MB of the companion star.

\subsubsection*{Magnetic Braking}
\label{sec:MB}
 In close binaries ($\Porb < 3~$d), the angular momentum lost by the companion is replenished at the expense of the orbital angular momentum of the binary through tidal interactions \citep{Tauris2023PhysicsSources}. As a result, the donor star is forced to fill its RL and initiate mass transfer. The AML by MB is conventionally calculated as

 \begin{equation}
    \frac {{\dot J}_{\rm{MB,Sk}}}{{J}_{\rm{orb}}} = - 3.8 \times 10^{-30}~f~\frac {{R_{\odot}}^{4}~({{R_2}/{R_{\odot}})^{\rm{\gamma_{MB}}}}~{G({M_1}+{M_2})^2}}{a^5~M_1}~{\text s}^{-1}
    \label{eq:MB1}
\end{equation}
 
where $R_2$ is the donor star radius, $\gamamb$ is the MB index ($\gamamb \simeq 2-5$), and $f = 0.73$ \citep{Skumanich1972TimeDepletion}. Equation (\ref{eq:MB1}) is also known as the Skumanich law \citep{Rappaport1983ABraking.}. The conventional MB prescription underestimates the $\Mdot_2$ of LMXBs by an order of magnitude \citep{Podsiadlowski2002EvolutionaryBinaries,Pavlovskii2016MassX-1}. To remedy this discrepancy, a modified MB model has been proposed by \cite{Pavlovskii2016MassX-1} and was later extended in the works of \cite{Van2019Low-massPrescription} (hereafter VIH19) and \cite{Van2019EvolvingBraking, Van2021ConstrainingBraking}. In the Skumanich law, the AML due to MB depends on $M_2, R_2$, and the rotational rate of the companion star $\Omega_2$ as: $\dot{J}_{{\rm{MB,Sk}}} \propto  ~M_{2}~R_{2}^{\rm{\gamma_{MB}}}~\Omega_2^3$. Assuming isotropic isothermal winds, a radial magnetic field and a rotational boost of the magnetic field, VIH19 obtained a MB torque that scales the same as the Skumanich law. The VIH19 MB model considers an additional scaling of MB with the wind mass-loss rate of the companion, together with a scaling of the magnetic field strength with the turnover time of convective eddies. The general form of the VIH19 MB law is given as 

\begin{equation}
\dot{J}_{\mathrm{MB}, \text { boost }}=\dot{J}_{\mathrm{MB}, \mathrm{Sk}}\left(\frac{\Omega_2}{\Omega_{\odot}}\right)^{\rm{\beta_{\rm boost}}}\left(\frac{\tau_{\text {conv }}}{\tau_{\odot, \text { conv }}}\right)^{\xi_{\rm boost}}\left(\frac{\dot{M}_{\mathrm{2,W}}}{\dot{\mathrm{M}}_{\odot, \mathrm{W}}}\right)^{\rm \alpha_{\rm boost}}
\end{equation}

where $\Omega_{\odot} \approx 3\times 10^{-6}~$s$^{-1}$ is the angular frequency of the Sun, $\tau_{\text {conv}}$ is the turnover time of convective eddies, $\tau_{\odot, \text { conv }} = 2.8 \times 10^{6} \mathrm{~s}$ \citep[see][for details]{Van2019Low-massPrescription}, $ \dot{M}_{ \mathrm{2,W}}=2.52 \times 10^{13} \left(\frac{R_{2}}{R_{\odot}}\right)\left(\frac{L_{2}}{L_{\odot}}\right)\left(\frac{M_{\odot}}{M_{2}}\right)~ \rm{g}~ s^{-1}$ is the wind mass-loss rate of the companion, and $L_2$ is the luminosity of the companion \citep{Reimers1975CircumstellarGiants.}. Depending on the values of $(\beta_{\rm boost,\xi_{\rm boost},\alpha_{\rm boost})}$, there are four cases considered in the VIH19 MB law: default   $(0,0,0)$, convection-boosted ($\tau$- boosted) $(0,2,0)$, intermediate $(0,2,1)$ and wind-boosted $(2,4,1)$. Using \mesa, VIH19 simulated $2136$ binaries for each of the four cases to compare the impact of these prescriptions against known LMXB populations. While the intermediate MB model is better at reproducing the properties of observed LMXBs, the $\tau$-boosted MB also yields results in good agreement with observations. Independent support for the $\tau$-boosted model was later provided by \citet{Deng2021EvolutionPrescriptions}, who combined detailed binary evolution with binary population synthesis across both the LMXB and RMSP stages. Motivated by these results, we adopt the $\tau$-boosted model, which can be written as

\begin{equation}
    \frac {{\dot J}_{\rm MB}}{{J}_{\rm{orb}}} =\frac {{\dot J}_{\rm{MB,Sk}}}{{J}_{\rm{orb}}} \left(\frac{\tau_{\text {conv }}}{\tau_{\odot, \text { conv }}}\right)^{2}
    \label{eq:MB2}
\end{equation}

\newpage
\subsection{Physical assumptions}

There are $\sim 190$ LMXBs hosting NSs that have been observed with $\Lx$ varying between $10^{31}$ and $10^{38}~$\ergpers. Sources with $\Lx < 10^{35}~$\ergpers~are mostly transient and in quiescence, while those with $10^{35} < \Lx < 10^{38}~$\ergpers~are either transient sources in their outburst states or persistent sources \citep[for details, see][]{Bahramian2022}. There is a critical mass accretion rate, $\Mdotcrit$ that depends on $\Porb$, below which the accretion disc becomes viscously unstable. Below this threshold ($\Mdot_1 < \Mdotcrit$), the system becomes a transient source that shows cyclic outbursts with durations from a few weeks to several months and recurrence times varying from a few months to several years, indicating that the transient sources are predominantly in their quiescent states \citep[for details of disc instability model see e.g.][]{Dubus1999,Lasota2001TheTransients}. All known AMXPs are transient sources. Among these systems, some sources like HETE J1900.1-2455 and MAXI J0911-655 exhibited extended outburst phases lasting up to $7$ years. We adopt the $\Mdotcrit$ as estimated by \citet{Dubus1999}, given by the relation

\begin{equation}
    \dot{M}_{\rm{crit}}=~2.36\times 10^{15}\left(\frac{M_{\rm{1}}}{1.4~\rm{M}_{\odot}}\right)^{1/2}\left(\frac{M_{\rm{2}}}{\rm{M}_{\odot}}\right)^{-1/5}\left(\frac{\Porb}{\rm{~h}}\right)^{7 / 5} \rm{g}~ s^{-1}
\end{equation}

The mass accretion rate on to the NS, $\Mdot_1$, is assumed to be Eddington limited which is given in VIH19 as

\begin{equation}
\dot{M}_{\mathrm{Edd}} \approx 2.14 \times 10^{18} \frac{1}{1+X}~ \rm{g}~ s^{-1}
\end{equation}
 where $X$ is the hydrogen mass fraction in the material transferred from the donor.

\begin{figure*}
    \centering
    \includegraphics[width=0.97\linewidth]{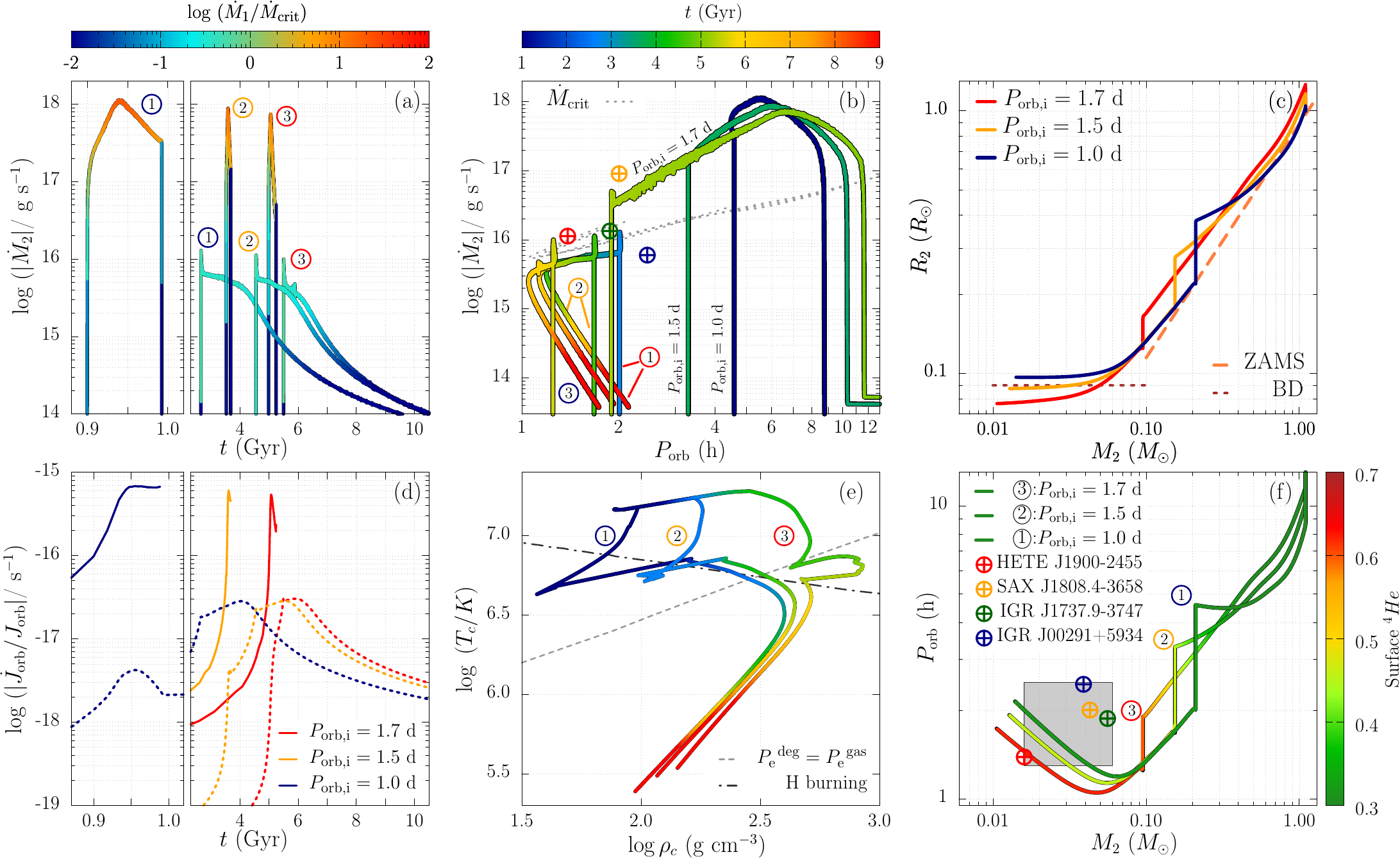}
    \caption{ Illustrative model tracks representing LMXBs evolving along the BD-track obtained with $\Porbi =$ $1.0$ (curve 1), $1.5$ (curve 2), and $1.7$ (curve 3) days. For the three model sources, $M_{\mathrm{1,i}} = 1.4 ~\Msun$,~$M_{\mathrm{2,i}} = 1.1 ~\Msun$, and $\gamamb = 3$. (a) and (b) show the evolution of $\dot{M}_2$ as a function of time and $\Porb$, respectively. The colour bar in (a) and (b) indicates the $\dot{M}_{\mathrm{1}/ \dot{M}_{\mathrm{crit}}}$ to distinguish the persistent and transient phases of the evolution and the time, respectively. The dashed curve in (b) is $\dot{M}_{\mathrm{crit}}$. The symbols "$\bigoplus$" show AMXPs with BD companions given in Table \ref{TableAMXP}. (c) Evolution of $R_2$ as a function of $M_2$. (d) Evolution of the AML rates $\dot{J}_{\mathrm{MB}}$ (solid lines) and $\dot{J}_{\mathrm{GR}}$ (dashed lines). (e) Evolution of the companion in the  temperature-central density $(T_{\rm c} - \rho_{\rm c})$ plane. The dashed line represents the border between the degenerate matter (below the line) and the non-degenerate matter (above the line) and the dot-dashed line is the critical line for the hydrogen burning. (f) Evolution of $\Porb$ as a function of $M_2$ with the colour bar indicating the donor's surface abundance in helium. The grey area indicates the regions in which this subgroup of AMXPs are observed. }
    \label{fig:BDtrack}
\end{figure*}

\subsection{Input physics and initial conditions}
To obtain the evolutionary tracks of AMXPs, we used the 1D stellar evolution code \mesa, version 11701 \citep[][]{Paxton2011ModulesMESA,Paxton2013ModulesStars,Paxton2015ModulesExplosions,Paxton2018ModulesExplosions,Paxton2019ModulesConservation}. In all the simulations, the NS is considered as a point mass evolving with a zero-age main sequence (ZAMS) companion with metallicity $Z = 0.02$. We treat convection according to the Schwarzschild criterion with a mixing length parameter of $\alpha_{\rm{MLT}} = 2.0$ and account for thermohaline mixing with a coefficient $\alpha_{\rm{th}} = 1.0$. We allow wind mass-loss of the donor via a Reimers-type wind. We chose $M_{\rm{1,i}}= 1.4 \Msun$ and $M_{\rm{2,i}} = 1.1 \Msun$ as canonical values where $M_{\rm{1,i}}$ is the initial NS mass, and $M_{\rm{2,i}}$ is the initial donor mass. Since we are interested in evolutionary tracks evolving to AMXPs (convergent systems), we chose initial orbital period values $\Porbi < 3~$d. We choose $\beta = 0.5$ and $\gamamb = 3$ as canonical values and discuss the impact of different values of $\beta$ and $\gamamb$ in Section \ref{subsec:beta-gamma_binary}. We note that our results are not  very sensitive to $q, \gamamb$, and $\beta$. With different values of these parameters, similar evolutionary tracks can be obtained with small adjustments to $\Porbi$ (see Section \ref{subsec:beta-gamma_binary}).

\subsection{The evolutionary tracks of AMXPs }
\label{subsec:The evolutionary tracks of AMXPs}

Here we present the results of our simulations with the \mesa~code that can reproduce the three evolutionary tracks of AMXPs. We obtain these illustrative evolutionary tracks, changing only the $\Porbi$ parameter of the model binaries in the ranges of $1.0 - 2.0~$d for BD-tracks, $1.7 - 3.0~$d for UCXB-tracks, and $1.2-3.0~$d for MS-tracks. All other parameters described above (given in the figure captions) are held constant throughout these simulations. Each track is evaluated against four key criteria that must be met simultaneously within a Hubble time to be counted as an evolutionary path for AMXPs: (i) The system must enter a transient regime towards the end of its evolution; (ii) the final values of $M_2$ and $\Porb$ must fall within the observed ranges specific to each group of AMXPs; (iii) the surface composition should be relatively hydrogen-rich for AMXPs with MS and BD companions, and helium-rich for those with He WD companions; (iv) the mass and radius of the donor star must correspond to those of the observed donor types in  AMXPs. 

\begin{figure*}
    \centering
    \includegraphics[width=0.98\linewidth]{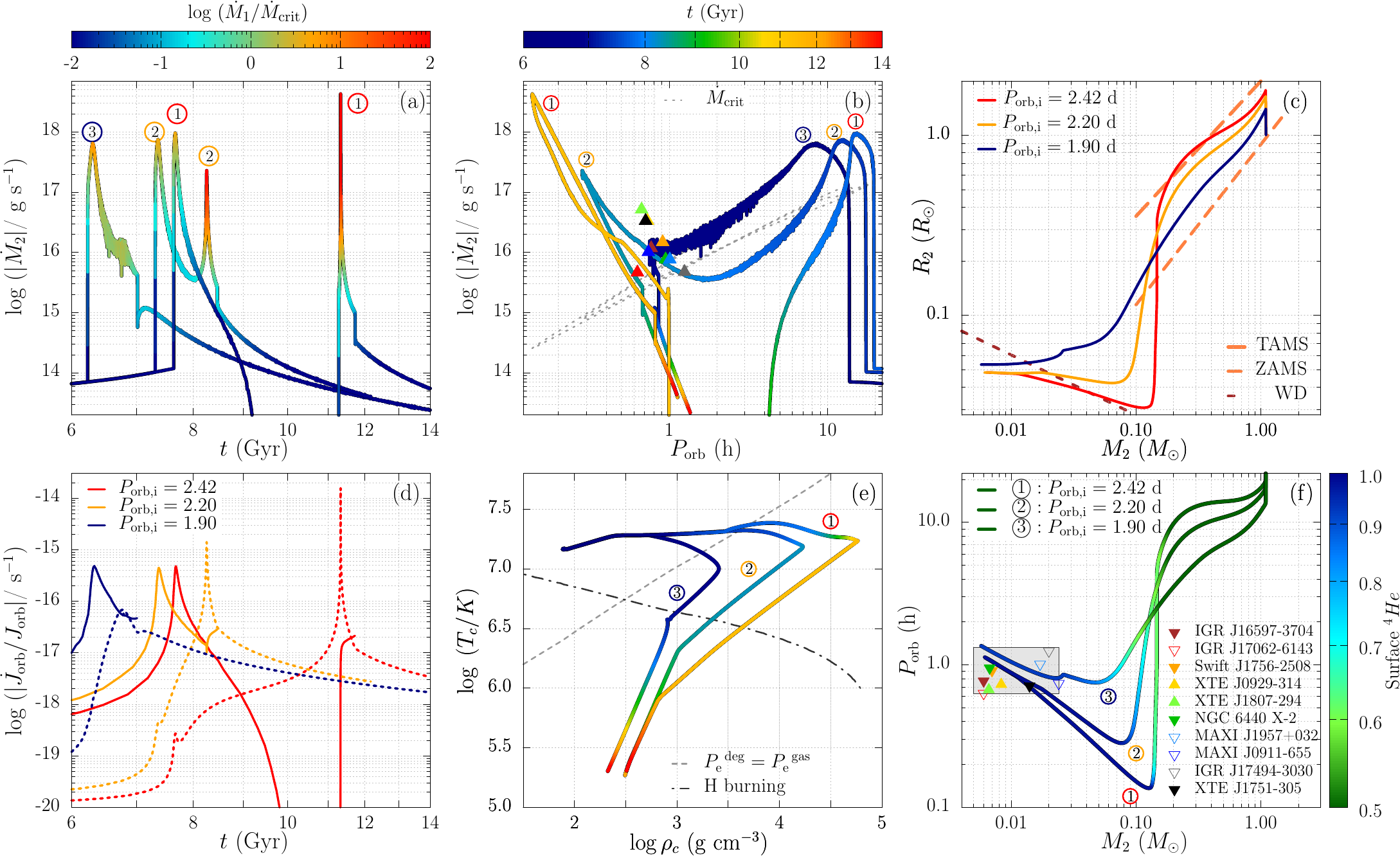}
    \caption{Same as Fig. \ref{fig:BDtrack}. These illustrative model tracks that can represent typical UCXB evolutions are obtained with $\Porbi =$ $2.42~$ d (curve 1), $2.20~$ d (curve 2), and $1.90~$ d (curve 3). The "$\blacktriangledown$" signs show observed AMXPs with He WD companions, the "$\blacktriangle$" signs are observed AMXPs with C/O WD companions and the "$\bigtriangledown$" signs represent observed AMXPs in UCXBs with some uncertainty in the companion type.}
    \label{fig:UCXBtrack}
\end{figure*}

The results for the model binaries are illustrated in Figs \ref{fig:BDtrack}–\ref{fig:MStrack}, showing the evolution of binary properties ($\dot{J}_{\mathrm{orb}}/J_{\mathrm{orb}}$ and $\Porb$) and donor properties ($M_2, R_2$, and $\Mdot_2$). To assess the validity of our evolutionary tracks for different groups of AMXPs, we plot these tracks alongside actual AMXPs on the $\Porb - \Mdot_2$ and $M_2 - \Porb$ planes (panels b and e in Figs \ref{fig:BDtrack}–\ref{fig:MStrack}). We summarize the details for each track in the following.

\subsubsection{The BD-track}
\label{subsec:BDtrack}

AMXPs with BD companions have $1~$h $< \Porb < 3~$h and $M_2 \simeq 0.01 \Msun$. For a MS donor to evolve to a BD star within Hubble time, the mass transfer must be initiated on a time-scale shorter than the nuclear time-scale, $\tau_{\rm{nuc}}$, before the companion burns a significant fraction of its core hydrogen to helium. This condition is satisfied for $0.5~$d $ < \Porbi < 1.7~$d. It is seen in Fig. \ref{fig:BDtrack}a,d that the three model binaries have the same $\Mdot_2 (t)$ and $\dot{J}_{\rm orb}(t)$ morphologies.

For $\Porbi = 1.0~$d, the orbital separation decreases due to AML via MB. In this phase, the MB strength reaches a maximum and the RL radius cannot increase fast enough and the mass transfer via RLOF is initiated when $\Porb =  {\Prlof} \simeq 9~$h, {where ${\Prlof}$ is the value of $\Porb$ at the onset of RLOF}, and $\Mdot_2 \sim 10^{18}~$\gpers~($\Mdot_2 >  \dot{M}_{\rm{crit}}$, the system is persistently accreting). The initial $\Mdot_2$ depends on the chosen MB model and the value of ${\Prlof}$, which also depends on the chosen MB model \citep[see the discussion in][] {Deng2021EvolutionPrescriptions, Yang2024TheBinaries}. Indeed, for similar parameters, initial $\Mdot_2$ is a few times greater with the conventional Skumanich MB model. As the strength of the MB torques decreases, $\Mdot_2$ decreases below $\Mdotcrit$ and the system enters the transient regime (see Fig. \ref{fig:BDtrack}a). The donor star becomes fully convective during the early stage of mass transfer inactivating the MB torques. At this stage ($\Porb < 3~$h), the donor has transferred nearly $90 \%$ of its mass. During the RLOF, the donor shrinks due to the mass loss (see Fig. \ref{fig:BDtrack}c). The RL radius cannot readjust fast enough within the thermal time-scale and the RLOF terminates until the relatively weak GR torques reduce the orbital separation and initiate a second RLOF phase when $t \sim 3~$Gyr. The correlation between $\Mdot_2$ and the torques acting on the binary can be seen in Fig. \ref{fig:BDtrack}a,d. GR torques are dominant in close binary systems ($\Porb < 3~$ h), and strong enough to drive the system into another RLOF phase during which the system evolves as a transient source ($\Mdot_2 \sim  10^{15}~$\gpers $~< \Mdotcrit$). The donor becomes fully degenerate when it moves below the theoretical line that separates non-degenerate matter and degenerate matter at $t \sim 5~$Gyr (see dashed line in Fig. \ref{fig:BDtrack}e). The donor evolves towards the BD region when $M_2 \sim 0.05 \Msun$. The mass-radius relation of the donor corresponds to that of BD \citep[$R_2 \propto  0.1 M_2$ for $M_2 < 0.1 M_{\odot}$;][]{Zdziarski2016IGRStar} (see Fig. \ref{fig:BDtrack}c). When the evolution terminates at $t \sim 14~$Gyr, $\Porb \sim 2~$h, and $M_2 \sim 0.014~\Msun$. At this time, the surface helium abundance of the donor is  $\sim 32 \%$. This is due to the onset of RLOF during the early phases of nuclear burning. Consequently, the change in surface helium abundance is negligible compared to models with larger $\Porbi$, where the abundance increases significantly as the donor evolves (see Fig. \ref{fig:BDtrack}e,f).

For the other model binaries, ${\Prlof} \simeq 10$~h for $\Porbi = 1.5$~d and ${\Prlof} \simeq 12$~h for $\Porbi = 1.7$~d. We observe that for given $M_{\rm{1,i}}$ and $M_{\rm{2,i}}$ values, ${\Prlof}$ increases with increasing $\Porbi$, a measure of orbital separation, since the RL radius is proportional to the orbital separation \citep{Eggleton1983APPROXIMATIONSLOBES}. In each case, $\Mdot_2$ is similar but slightly smaller compared to the $\Mdot_2$ of the model with $\Porbi = 1.0$~d. This is due to the slight difference in strength of MB at the beginning of mass transfer, which decreases with $\Porbi$ (see Fig. \ref{fig:BDtrack}a,d). The duration of the initial RLOF increases with increasing $\Porbi$ ($\sim 0.2~$Gyr for $\Porbi = 1.5~$d and $\sim 0.3~$Gyr for $\Porbi = 1.7~$d). As a result, the final donor mass is also smaller for greater $\Porbi$ ($M_2 \simeq 0.012~ \Msun$ for $\Porbi = 1.5~$d and $M_2 \simeq 0.010~ \Msun$ for $\Porbi = 1.7~$d; see Fig. \ref{fig:BDtrack}c). The second RLOF is initiated at $t \sim 5~$Gyr for $\Porbi = 1.5~$d and $t \sim 6~$Gyr for $\Porbi = 1.7~$d. Nuclear burning of the donor also lasts longer for binaries with greater $\Porbi$. This leads to a higher surface helium abundance of the donor by the end of its evolution ($\sim 45 \%$ for $\Porbi = 1.5~$d and $\sim 60 \%$ for $\Porbi = 1.7~$d). 

We plotted the AMXPs with BD donors on the $M_2 - \Porb$ plane alongside three model binary systems with $\Porbi =$ $1.0$, $1.5$, and $1.7$ days, respectively (see Fig. \ref{fig:BDtrack}e). The observed $\Porb$ and $M_2$ of the source HETE J$1900.1-2455$ can be reproduced on the BD-track with $\Porbi = 1.7~$d, which is in agreement with observations. We could not find evolutionary tracks that match the observed properties of the remaining sources SAX J1808.4-3658, IGR J00291+5934, and MAXI J1957+032 on the $M_2 - \Porb$ plane. These sources have observed $\Porb$ values slightly larger than those predicted by our evolutionary tracks. Studies of the formation and evolution of spider pulsars indicate that the black widow pulsars share similarities with AMXPs with BD donors \citep{Chen2013FormationPulsars., Chen2017AnJ1808.43658}. Indeed, they have similar observed $M_2$ and $\Porb$, while black widows have relatively higher $\Porb$. This orbital expansion is likely caused by the irradiation of the companion by the NS as it spins down between mass transfer phases. The spin-down luminosity of the NS irradiates the companion, causing its evaporation through a stellar wind, which results in orbital expansion \citep{Chen2013FormationPulsars., Misra2025InvestigatingEvolution}. This irradiation effect is not addressed in our binary models, which is likely to be the reason our analysis could not accurately reproduce the properties of SAX J1808.4-3658, IGR J00291+5934, and MAXI J1957+032 on the $M_2 - \Porb$ plane.

We plot the same sources on the $\Porb - \Mdot_2$ plane together with three model binaries with $\Porbi =$ $1.0$, $1.5$, and $1.7$ days, respectively (see Fig. \ref{fig:BDtrack}b). The grey dashed lines correspond to $\Mdotcrit$. To plot the AMXPs with BD donors, we use $\Mdot_2 = \Mdot_{\rm avg}$, where $\Mdot_{\rm avg}$ corresponds to values seen in Table \ref{TableAMXP}. Here, $L_{\rm X, avg}$ represents the average peak luminosities of the sources observed during outbursts \citep{Heinke2025CatalogBinaries}. As seen in Fig. \ref{fig:BDtrack}b, the four sources remain above the $\Mdotcrit$ line (while IGR J00291+5934 is close to but below the line). Furthermore, our model tracks indicate that these sources are close to the onset of the second accretion phase. Most of these systems are likely to have completed their SU phase.

\subsubsection{The UCXB-track}
\label{subsec:UCXBtrack}

UCXBs are systems with observed $\Porb < 80~$min and an extremely low-mass, degenerate donor He/CO WD with $M_2 \simeq 0.001 \Msun$. Model binaries for UCXB-tracks can be obtained with $1.7~$d $< \Porbi < 3.0~$d (Fig. \ref{fig:UCXBtrack}). The model BD-tracks in Fig. \ref{fig:BDtrack} are sensitive to the chosen $\Porbi$ and show similar morphology. However, the UCXB model tracks we obtain, which are also sensitive to the chosen $\Porbi$, do not exhibit the same morphology. We distinguish three different cases, which are summarized below. 

The evolution of the model source with $\Porbi \simeq 2.42~$d represents Case 1 in Fig. \ref{fig:UCXBtrack}, the donor evolves on $\tau_{\rm{nuc}}$ and burns most of its core hydrogen before initiating mass transfer at around the terminal-age main sequence (TAMS) at $t \sim  7.5~$Gyr. The orbital separation decreases due to AML via MB, and the RLOF starts with a rate rapidly increasing to $\Mdot_2 \sim 10^{18}~$\gpers~when $\Porb \simeq 20~$h. During this stage, the MB torque is at its strongest level. With decreasing strength of MB torques, and thus $\Mdot_2$, the system enters a transient regime (see Fig. \ref{fig:UCXBtrack}). Around $\Porb \sim 4~$h, the donor evolves away from the TAMS. After $\sim 90 \%$ of its mass was transferred, the donor can no longer fill its RL, the RLOF phase terminates switching off the LMXB phase. $\Porb$ continues to decrease to $\sim 1~$h while both MB and GR torques are active. During this time interval that lasts $ \sim 2.0~$Gyr, the source is not detectable as an X-ray source while it could be observed as a RMSP \citep{Sengar2017NovelStars,Tauris2018DisentanglingLISA}. This time interval increases with increasing $\Porbi$ (e.g. $ \sim  3.5~$Gyr for $\Porbi \sim 2.5~$d).  With increasing strength of the GR torques, the orbital separation continues to decrease and finally initiates a second RLOF when $\Porb \sim 1$~h. The donor is now a fully degenerate WD (see Fig. \ref{fig:UCXBtrack}c,e), and is able to refill its RL. Since the radius of a degenerate WD increases with decreasing mass \citep[$R_{\rm{WD}} = 0.013~R_{\odot}~ ( {M_{\rm{WD}} /M_{\odot}})^{-1/3}$,][]{Sengar2017NovelStars}, the donor responds to mass loss by expanding. The resultant UCXB phase begins with $\Mdot_2 \sim \Mdot_{\rm Edd}$ with persistent accretion. The orbital separation continues to decrease until $\Porb$ reaches a minimum around $10~$min. The $\Mdot_2 - \Porb$ correlation during the UCXB has been studied by \cite{Sengar2017NovelStars}, who showed that, toward the end of the evolution, all UCXB-tracks follow a common \textit{declining} branch during which the orbit expands. Most of the remaining hydrogen envelope of the companion is accreted during the UCXB phase. The evolution ends with a transiently accreting NS, and a He WD donor with $M_2 \simeq 0.006 \Msun$ and surface helium abundance of $98\%$ (see Fig. \ref{fig:UCXBtrack}b,f).

For Case 2 ($\Porbi = 2.20~$d), the initial steps of the evolution (LMXB phase) are rather similar to Case 1, with ${\Prlof} \simeq 17~$h. With a relatively shorter orbital separation, the GR torques are comparable to the MB torques at an earlier stage of the LMXB phase. As a result, the system enters into a transient regime during which $\Mdot_2$ decreases and reaches a minimum $\sim 3 \times 10^{15}~$\gpers. As the strength of the GR torques increases, $\Mdot_2$ starts to increase, and the UCXB phase begins with persistent accretion (see Fig. \ref{fig:UCXBtrack}a,d). The evolution terminates similar to Case 1, when the surface helium abundance of the donor is about $98 \%$. Note that the detached phase encountered in Case 1, while the source could be observed as a RMSP, is not present in Case~2. 

Case 3 ($\Porbi = 1.90~$d) can be considered as an extreme case of the BD-track with a greater value of ${\Prlof} \simeq 14~$h. Unlike Case 1 and Case 2, MB torques become negligible when the donor becomes degenerate (see Fig. \ref{fig:UCXBtrack}d,e). Afterwards, RLOF is maintained by the GR torques. The system evolves in the transient regime for the remainder of its lifetime, which ends when the companion has $\sim 92 \%$ surface helium abundance. 

In Fig. \ref{fig:UCXBtrack}f, the AMXPs with He WD donors are plotted on the $M_2 - \Porb$ plane alongside three illustrative model binary systems with $\Porbi =$ $2.42$, $2.00$, and $1.90$ days. These model curves are consistent with the observed $M_2$ and $\Porb$ values of most AMXPs with He WD donors. However, XTE J0929-314 and XTE J1807-294, which host low-mass C/O WD companions, are likely to have evolved from intermediate mass X-ray binaries with $M_{\mathrm{2,i}} \geq 1.5~ \Msun$, a parameter range that was not explored in our analysis. The same sources and model tracks are also shown on the $\Porb - \Mdot_2$ plane in Fig. \ref{fig:UCXBtrack}b. These sources are typically located near the terminal stages of the model tracks. This indicates that the obtained model tracks are indeed representative of AMXPs with He WD companions.

\begin{figure*}
    \centering
    \includegraphics[width=0.97\linewidth]{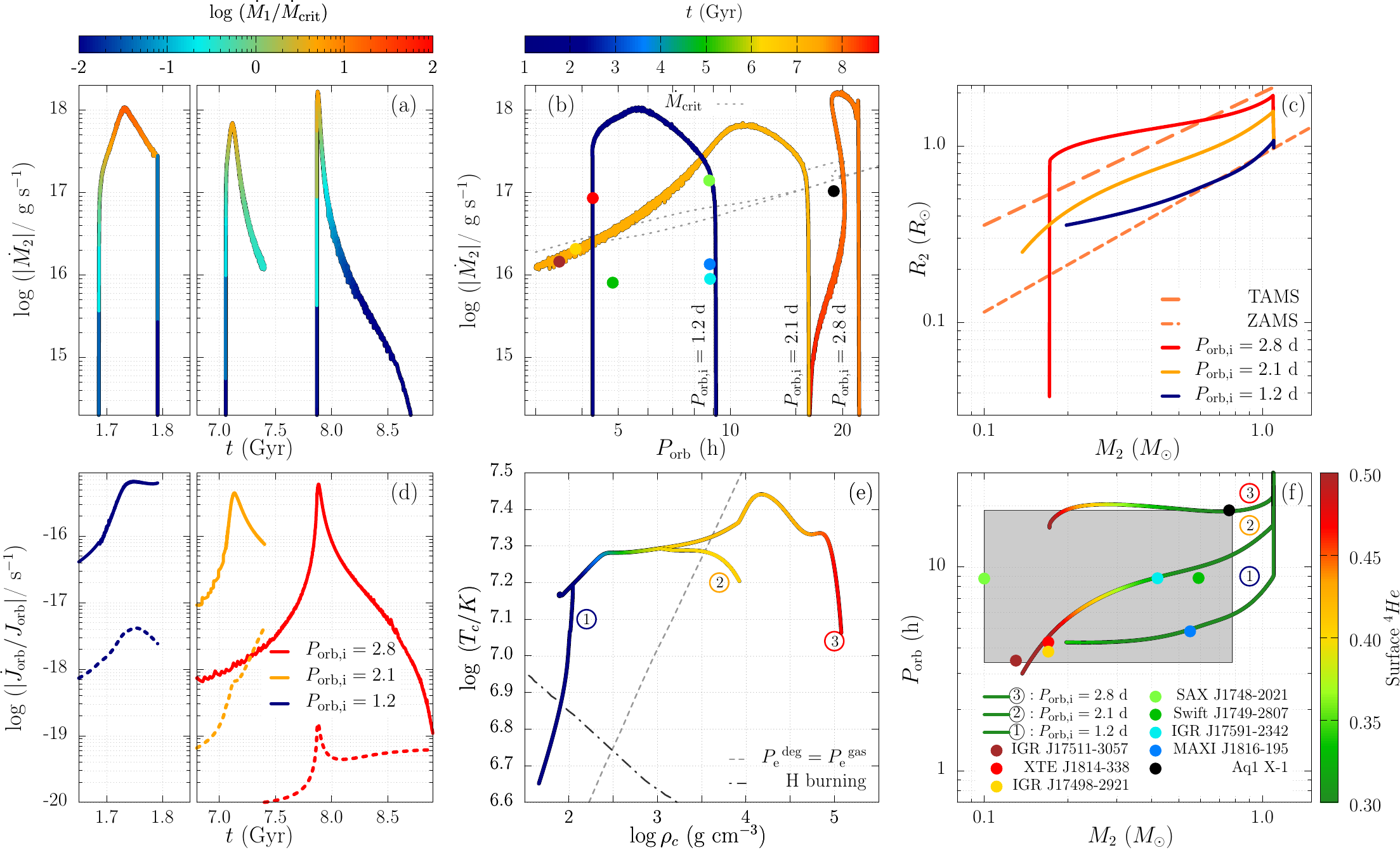}
    \caption{Same as Fig. \ref{fig:BDtrack}. These model curves that can represent the evolution of AMXPs with MS companion are obtained with $\Porbi =$ $2.8$, $2.1$, and $1.2$ days. The filled circles show observed AMXPs with MS companions.}
    \label{fig:MStrack}
\end{figure*}

\subsubsection{The MS-track}
\label{subsec:MStrack}

AMXPs with MS companions have $3~$h$< \Porb < 19~$h and $0.1~\Msun < M_2  < 0.5~\Msun$. Our results imply that these systems are likely to be observed at an early stage of their evolution on either BD or UCXB tracks. To illustrate this, we present one BD-track with $\Porbi~=~1.2~$d and two UCXB-tracks with $\Porbi~=~2.1~$d and $\Porbi~=~2.8~$d in Fig. \ref{fig:MStrack}. We run these simulations until $\Porb = 3~$h and/or $M_2 = 0.1~\Msun$.  

\newpage
The UCXB-tracks described in Section \ref{subsec:UCXBtrack} include two different long-term phases of accretion: an LMXB phase that is driven by MB torques during the early stage of the evolution and a UCXB phase that is driven by the GR torques, during the late stage of the evolution. For some systems, GR torques may not become strong enough to evolve into a UCXB phase. In such cases, the evolution terminates right after the LMXB phase. These "pre-matured" UCXB-tracks terminate with properties similar to those of AMXPs with evolved MS companions. For these evolutionary paths, GR torques remain weaker than MB torques by an order of magnitude as the MB torques decrease. The pre-matured UCXB-tracks can be obtained with $\Porbi$ values in the range of $2.5~$d $\lesssim \Porbi \lesssim 3.0~$d, which lead to convergent systems. 

For the model source shown in Fig. \ref{fig:MStrack} with $\Porbi \sim 67~$h ($2.8~$d), the donor star evolves on $\tau_{\rm{nuc}}$, and approaches the TAMS before the mass transfer is initiated at $t \sim 8~\mathrm{Gyr}$. The AML via MB gradually reduces the orbital separation, and RLOF starts when the $\Porb$ decreases to $\simeq 22~\mathrm{h}$. During this phase, the MB torques reach maximum strength, the resultant decrease in the size of the RL initiates a rapid mass loss at a rate of $\Mdot_2 \sim 10^{18}~$\gpers. Since the MB torques remain weak and the GR torques become even weaker, the orbital separation does not decrease significantly. Decreasing $\Mdot_2$ takes the system into the transient regime (see Fig. \ref{fig:MStrack}a,d). This occurs when the donor has transferred $\sim 80 \%$ of its mass, and $\Mdot_2$ decreases below $10^{16}~$\gpers. RLOF terminates at $t \sim 8.5~\mathrm{Gyr}$ while mass loss via wind continues with rates $ < 10^{14}~$\gpers, without a significant effect on the binary evolution. We neglect the contributions of winds to $\Mdot_1$ in our calculations. The evolution terminates with an evolved MS companion when $\Porb \sim 18$~h and $M_2 \sim 0.2 ~\Msun $ (see Fig. \ref{fig:MStrack}b,c,e). The observed properties of AMXPs with MS companions are shown in Fig. \ref{fig:MStrack}f together with the three models with $\Porbi =$ $1.2$, $2.1$, and $2.8$ days. The model tracks can simultaneously reproduce the observed properties ($M_2, \Mdot_2, \text{and}~ \Porb$; Fig. \ref{fig:MStrack}b,f) of these AMXPs (with the exception of SAX J1748-2021).

The BD, UCXB, and MS tracks represent distinct evolutionary paths governed by AML through MB and GR. While the BD and UCXB tracks undergo early or delayed mass transfer leading to degenerate donors, the MS-tracks seem to correspond to the LMXB phase of UCXB-tracks ending with evolved MS companions. Each of these tracks produces a different mass transfer history, which we will use to investigate the rotational evolution of AMXPs in Section \ref{Section3}.

\begin{figure*}
    \centering
    \includegraphics[width=0.95\linewidth]{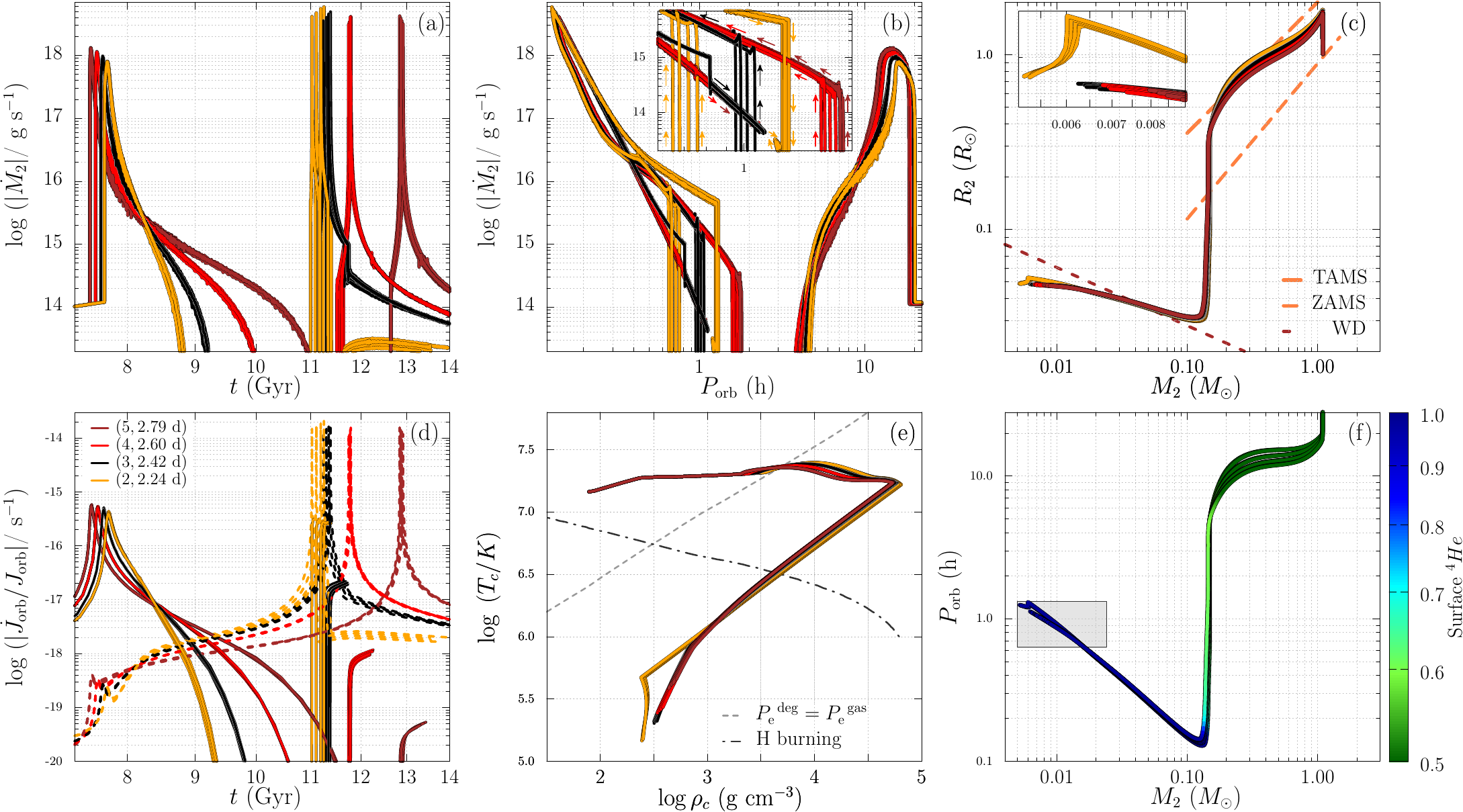}
    \caption{ Same as Fig. \ref{fig:UCXBtrack}. The effect of $\beta$ and $\gamamb$ is shown for the illustrative Case 1 of the UCXB-track through grid analysis. In panels (a)-(e), the values of $(\gamamb,\Porbi)$ pairs are $(2,2.24~\text{d})$ for orange curves, $(3,2.42~\text{d})$ for black curves, $(4,2.60~\text{d})$ for red curves, and $(5,2.79~\text{d})$ for brown curves. In the zoom-in of (b) arrows indicate the directions followed by the curves. The zoom-in in (c) allows a better reading of final $M_2$ attained by the model curves.}
    \label{fig:beta_gamma_effect}
\end{figure*}

\subsection{Effects of $\beta$ and $\gamamb$ on the binary evolution}
\label{subsec:beta-gamma_binary}

The key \mesa~parameters that affect the binary evolution are $\beta$, $\gamamb$, and $\Porbi$. To quantify the impact of these parameters, we performed a grid study in which we repeated the evolution of the illustrative Case~1 of the UCXB-track (see Section~\ref{subsec:UCXBtrack}). We chose Case 1 because it is an illustrative binary evolution, undergoing two accretion phases at rates close to the Eddington limit during the LMXB and UCXB phases. Consequently, the NS experiences two distinct spin-up episodes, separated by a detachment phase. This allows an in-depth analysis of how the key \mesa~parameters affects both the binary evolution and the rotational evolution of the NS. We constituted the grid for four $(\gamamb,\Porbi)$ pairs: $(2,2.24~\text{d})$, $(3,2.42~\text{d})$, $(4,2.60~\text{d})$, and
$(5,2.79~\text{d})$, and for $\beta = 0.2,\,0.4,\,0.6,$ and $0.8$. For each different $\gamamb$ value, we adjusted $\Porbi$ by $\sim 7\%$ in order to produce the same qualitative evolutionary sequence and enable a comparison of the resulting $\Mdot_2$ histories. Fig.~\ref{fig:beta_gamma_effect} shows the binary evolution of the $16$ binary models, all of which satisfy the four criteria outlined in Section~\ref{subsec:The evolutionary tracks of AMXPs}. In all cases, the donor ultimately acquires the characteristic properties of a He WD (Fig.~\ref{fig:beta_gamma_effect}c,e,f). The final donor masses obtained for different $\gamamb$ values ($\sim 5.2 \times 10^{-3}\,\Msun$, $\sim 6.2 \times 10^{-3}\,\Msun$, $\sim 6.8 \times 10^{-3}\,\Msun$, and $\sim 7.7 \times 10^{-3}\,\Msun$) lie within  $\sim 15\%$  of the value obtained for Case 1 (see the zoom-in in Fig. \ref{fig:beta_gamma_effect}c).

 The most significant differences across the grid are seen in the $\Mdot_2$ histories (Fig.~\ref{fig:beta_gamma_effect}a). While the detachment phase lasts $\sim 2$ Gyr in all $16$ models, the duration of the LMXB phase increases systematically with increasing $\gamamb$. This behaviour reflects the slower decay of MB torques for larger $\gamamb$ values (Fig.~\ref{fig:beta_gamma_effect}d; see also \citealt{Rappaport1983ABraking.}). Although MB torques are generally negligible compared to GR torques during the UCXB phase, they remain important for $\gamamb = 2$ and $3$. In particular, for $\gamamb = 2$, the system operates close to a balance between MB and GR torques, leading $\Mdot_2$ to decrease below $ 1 \x 10^{11}~$\gpers. The mass transfer resumes within $\simeq 2 \x 10^{-2}~$Gyr with $\Mdot_2 \sim 10^{13}~$\gpers. As a result, the corresponding evolutionary tracks do not terminate along the same branch as those obtained for $\gamamb \geq 3$ (see the zoom-in in Fig.~\ref{fig:beta_gamma_effect}b). For a given ($\gamamb, \Porbi$) pair, variations in $\beta$ primarily affect $\Mdot_2$, with lower (higher) $\beta$ values yielding higher (lower) {rates of mass transfer into the RL of the NS}.

The results of a similar grid analysis for the BD and MS tracks show that for different $\gamamb$ values, evolutionary curves with similar morphologies can be obtained by adjusting $\Porbi$ by a small percentage depending on $M_{\rm{2,i}}$. For instance, for a given $\gamamb$ and $M_{\rm{2,i}}$ values of $1.1~\Msun$ and $1.3~\Msun$, we obtain similar model curves with $\sim 7\%$ and $\sim 13\%$ adjustments in $\Porbi$, respectively. For given $\Porbi$, different $\gamamb$ values lead to qualitatively different evolutions. Models with $\gamamb = 3$, $4$, and $5$ robustly remain on the same evolutionary track, differing mainly in the details of $\Mdot_2$ history. In contrast, evolution with $\gamamb = 2$ is very  sensitive to $\Porbi$ such that small changes in $\Porbi$ could cause a transition to a different evolutionary track. This behaviour is reminiscent of the extreme fine-tuning problem identified by \citet{Istrate2014ThePeriod} for low $\gamamb$.

Overall, this grid study demonstrates that the $\Mdot_2$ histories are more sensitive to $\gamamb$ than to $\beta$. The effect of these parameters is similar for the BD and MS tracks as well. We will also use the results of this grid study to investigate the effect of $\beta$ and ($\gamamb,\Porbi$) pairs on the NS rotational evolution in Section \ref{subsec:beta-gamma_rotation}.

\section{Rotational evolution of AMXPs} 
\label{Section3}

\subsection{The Model}
\label{subsec:The model}

\begin{figure}
    \centering
    \includegraphics[width=0.85\columnwidth]{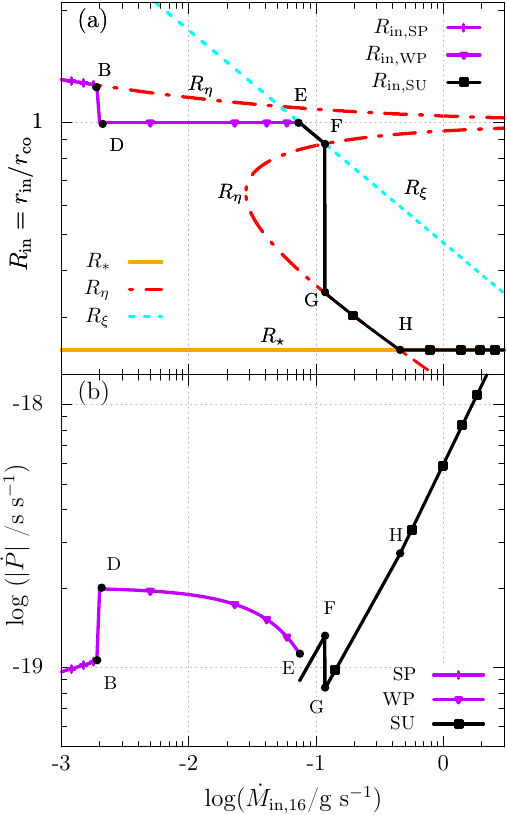}
    \caption{Variations of $\Rin$ (a) and $\Pdot$ (b) with $\Mdotin$ in the SP (solid curve with a plus sign), WP (solid curve with filled triangles) and SU phases (solid curve with filled squares). For this illustrative source, we take $r_{\star} = 1.2 \times 10^6~$cm, $\DeltaR = 0.2$, $\eta = 1.0$, $\xi = 0.5$, $B = \mu/r_{\star}^{3} = 1 \times 10^8$~G, and $P = 10$~ms (see the text for details). In panel (a), the dot-dashed curves and dashed line represent $\Reta (\Mdotin)$ and $\Rxi (\Mdotin)$, respectively, and the solid line shows, $R_{\star}$. The points with letters in panel (b) indicate the same transitions shown in panel (a) with the SP phase beyond B, the WP phase between D - E and the SU phase between points E - H.}
    \label{fig:Rin_Mdotin}
\end{figure}

Here we briefly describe how the rotational phases and the inner disc properties of a source change with gradually increasing $\Mdotin$ using an illustrative model source seen in Fig. \ref{fig:Rin_Mdotin} \citep[for details, see][]{Ertan2021}. In the model, there are three basic rotational phases: the strong propeller (SP), the weak propeller (WP), and the spin-up phase (SU). At the lowest $\Mdotin$ rates (lower than the rate corresponding to point B in Fig. \ref{fig:Rin_Mdotin}), the system is in the SP phase with $\rin > r_1 =1.26~\rco$. Here, $\rco = (GM_1/ \Ostar^2)^{1/3}$ is the co-rotation radius, and $\Ostar$ is the angular spin frequency of the NS. In this phase, all the inflowing mass is thrown out of the system from the narrow inner boundary of the disc. 

The maximum inner disc radius, $\rinmax$, at which the SP mechanism is sustainable is obtained from

\begin{equation}	
	\Rinmax^{25/8}~|1 - \Rinmax^{-3/2}| \simeq 0.22~\alpha_{\rm visc,-1}^{2/5}~M_{1.4}^{-7/6}~\Mdot_\mathrm{in,16}^{-7/20}~\mu_{26}~P^{-13/12}_\mathrm{ms}
        \label{eq:Rin}
\end{equation}

where $\Rinmax=\rinmax/\rco$, $M_{1.4} = (M_1 / 1.4 \Msun )$, $\Mdot_\mathrm{in,16} = \Mdotin / (10^{16}$~\gpers), $\mu_{26} = \mu / (10^{26}$~G~cm$^3$),  $\mu$ is the magnetic dipole moment of the NS, $\alpha_{\rm visc,-1} = (\alpha_{\rm visc} /0.1)$ 
is the kinematic viscosity parameter \citep{Shakura1973BlackAppearances}, and $P_\mathrm{ms}$ is the spin period in milliseconds \citep{Ertan2017TheStars}. We also define the radii $R_{\star} = r_{\star}/\rco$ where $r_{\star}$ is the NS radius, $\Reta = \eta \Rinmax = \eta~\reta/\rco$ with $\eta\lesssim 1$, $\Rxi = \xi \RA = \xi~\rA/\rco$ with $\xi \sim 0.5 -1$ \citep{Ghosh1979ACCRETION1}. The variations of $\Reta$ (dot-dashed curves) obtained from the solution of equation (\ref{eq:Rin}) and $\Rxi$ (dashed line) with $\Mdotin$ are shown in Fig. \ref{fig:Rin_Mdotin}a for an illustrative model source with $P = 5~$ms and $\mu_{26} = 10^{26}$~G cm$^{3}$.

For higher $\Mdotin$, if $\rin$ is instantaneously between $\rco$ and $r_1$ the matter expelled from the inner disc returns back to the disc at larger radii causing a pile-up, which pushes the inner disc inwards down to $\rin = \rco $ (B - D). This switches on the accretion on to the NS taking the system into the WP phase. For a large range of $\Mdotin$, the system can remain in the WP phase with $\rin = \rco$ while the spin-down torques dominate the spin-up torques produced by accretion (D - E) except when the star is close to the WP/SU transition. The inner disc can penetrate into $\rco$ when the viscous stresses dominate the magnetic stresses around $\rco$. This happens when $\Mdotin$ exceeds the rate corresponding approximately to $\rin = \rxi = \xi~ \rA$. The transition from the WP phase to the SU phase (torque reversal) also takes place at this stage. Beyond this critical rate, $\rin $ tracks $\rxi$ for a narrow $\Mdotin$ range (E - F) up to the point on the $\reta$ solution, which is also calculated from equation \ref{eq:Rin}. For  $\Rin < 1$ ($\rin <\rco$) the solution of equation (\ref{eq:Rin}) is double valued with an unstable upper branch and a stable lower branch \citep[see][for details]{Ertan2021}. For the $\Mdotin$ rate at point F on the unstable branch, the inner disc propagates inwards, opening the field lines down to the radius corresponding to point G on the lower stable branch of $\reta$ solution. With further increase in $\Mdotin$, $\rin$ decreases tracking the lower branch of $\reta$ until $\rin = r_{\star}$. Depending on the critical $P$ and $B$ values, in some cases the inner disc propagates directly on to the NS since $r_{\star} > \reta$ for the $\Mdotin$ corresponding to this transition \citep[for details, see][]{Ertan2021}.

The total torque acting on the NS can be written as:

\begin{equation}
	\Gamma =  ~ \sqrt{G M_1} \rin ~\Mdotstar - ~\frac {\mu^2}{\rin^3} \left(\frac{\Delta r}{\rin}\right) ~ -\frac{2}{3} \frac{\mu^2 \Ostar^3}{c^3}
\label{eq:torques}
\end{equation}

In this equation, the first term is the spin-up torque, $\Gamma_{\rm acc}$, which is associated with the mass accretion from the inner disc on to the NS \citep{Pringle1972AccretionSources}. Here, $\Mdotstar$ is the accretion rate on to the star. The second term is a spin-down torque, $\Gamma_{\rm D}$, which arises from the interaction between the inner disc and the field lines within the boundary layer with radial width $\Delta r$. The third term is the magnetic dipole torque, $\Gamma_{\rm dip}$, also a spin-down torque. The $\Pdot$ and $\rin$ variations are seen in Fig. \ref{fig:Rin_Mdotin}.

The total torques acting on the NS during each rotational phase are given as 

\begin{equation}
\left\{
\begin{aligned}
    \Gamma &= \Gamma_{\rm acc}  +  \Gamma_{\rm dip} & (\text{SU; } \rin < \rco) \\
    \Gamma &= \Gamma_{\rm acc} + \Gamma_{\rm D} + \Gamma_{\rm dip} & (\text{WP; } \rin = \rco) \\
    \Gamma &= \Gamma_{\rm D} + \Gamma_{\rm dip} & (\text{SP; } \rin > r_1 )
\end{aligned}
\right.
\end{equation}

The model does not address the evaporation of the inner disc due to thermal instabilities at very low $\Mdotin$ levels \citep{Frank2002AccretionEdition}.

\begin{figure*}
    \centering
    \includegraphics[width=0.9\linewidth]{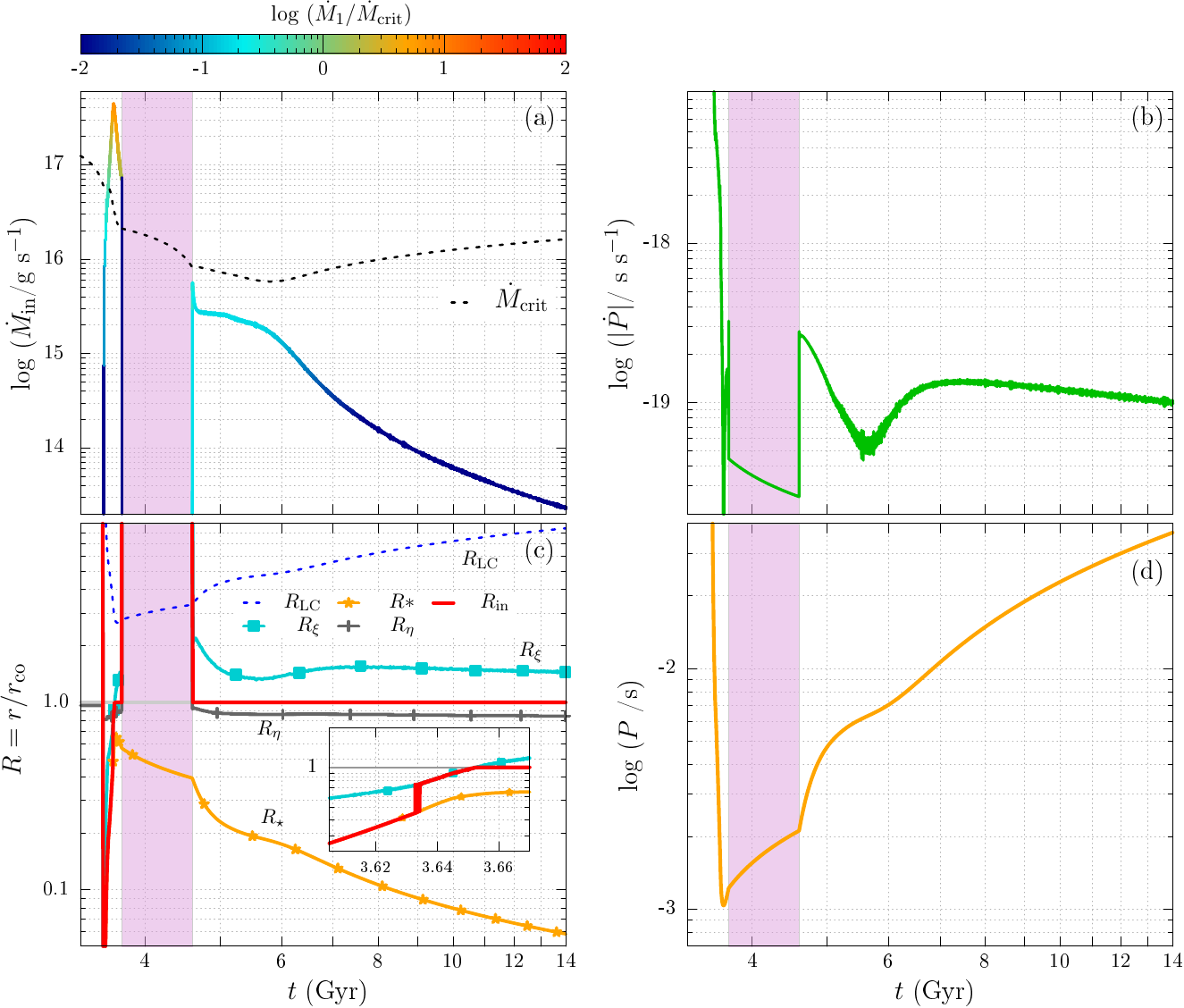}
    \caption{ Evolution of the rotational properties of a NS evolving on the BD-track shown in Fig. \ref{fig:BDtrack}. For all the curves, $r_{\star} = 1.2 \times 10^6~$cm, $\DeltaR = 0.2$, $\eta = 0.8$, $\xi = 0.8$, $B = \mu/R^{3} = 2 \times 10^8$~G, $P_{\rm i} = 100~$s, and $\Porbi = 1.5~$d. (a) Evolution of $\Mdotin$ with the colour bar indicating the ratio of $\dot{M}_{\mathrm{1}}/ \dot{M}_{\mathrm{crit}}$ and the dotted curve, $\Mdotcrit$. (b) Evolution of $\Pdot$. (c) Evolution of $\Rin$ (solid curve), $R_{\rm LC}$ (dashed curve), $\Rxi$ (solid curve with squares), $\Reta$ (solid curve with plus signs), and $R_{\star}$ (solid curve with stars). (d) Evolution of $P$. The shaded area indicate the RMSP phase.}
    \label{fig:BDtrack_evl}
\end{figure*}

\subsection{Numerical calculations}

Using \mesa, we have studied the binary evolution for each of the evolutionary tracks and obtained their respective $\Mdot_2$, $\Mdot_1$, and $M_1$ evolution. For the model calculations, we take  $r_{\star} = 12~$km, the moment of inertia $I \simeq 2~M_1~r_{\star}^2/5$, $\xi = 0.8$, $\eta = 0.8$, and $\DeltaR = 0.2$ for all the evolutionary tracks, which gave reasonable results in the earlier applications \citep{Gencali2022, Niang2024OnBinaries} and discuss the effects of the model parameters on the rotational evolution of the NS in Section \ref{subsec: Model parameters}. 

 \begin{figure*}
    \centering
    \includegraphics[width=0.9\linewidth]{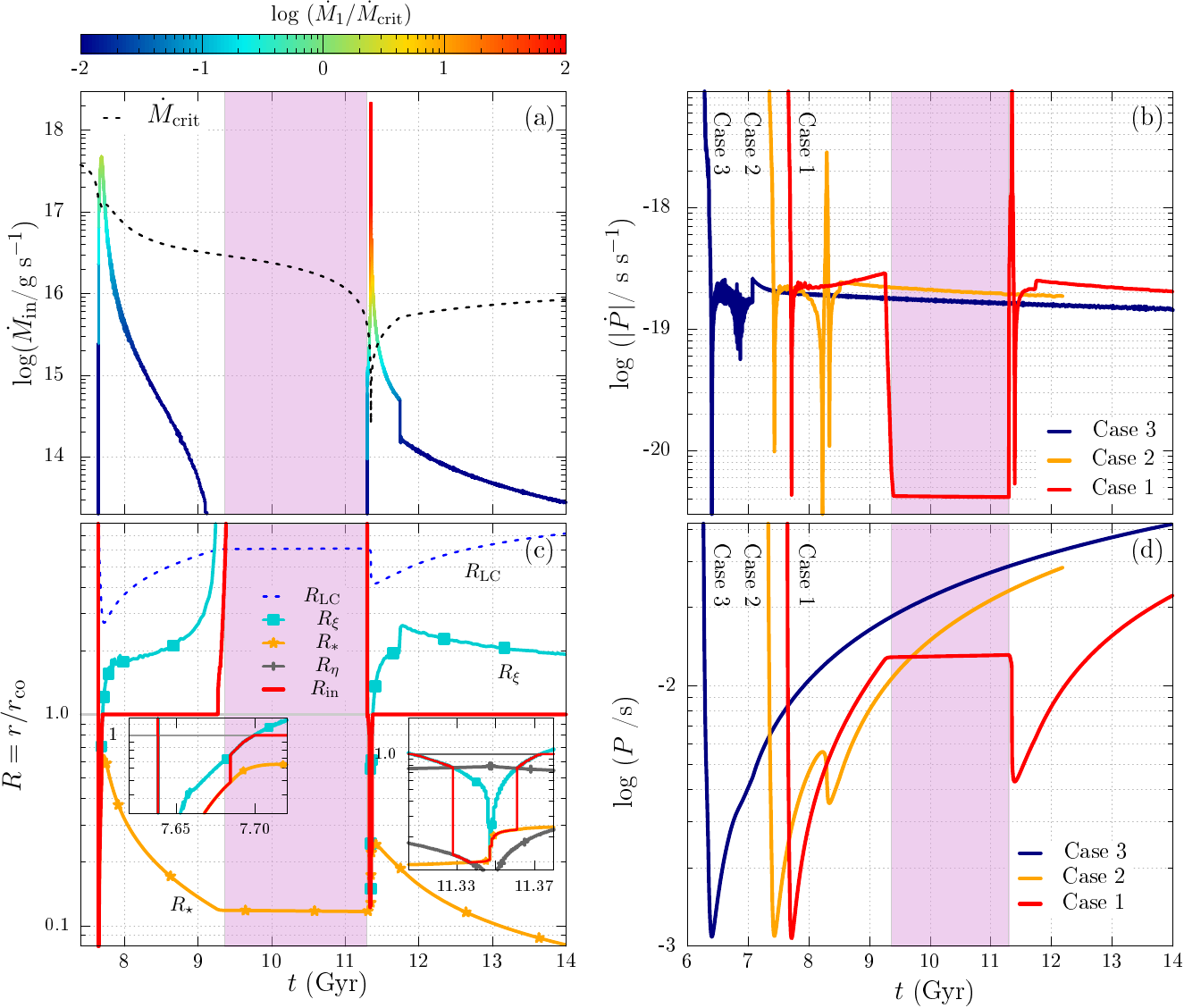}
    \caption{Same as Fig. \ref{fig:BDtrack_evl} for a NS evolving on a UCXB-track with $\Porbi = 2.42~$d (Case 1), $\Porbi = 2.00~$d (Case 2), and $\Porbi = 1.90~$d (Case 3).}
    \label{fig:UCXBtrack_evl}
\end{figure*}

 The magnetic field of recycled pulsars is estimated to decay during the early phases of the long-term evolution of LMXBs. The field decays rapidly via accretion-induced ohmic dissipation, on a time-scale much shorter than the time-scale of the spin-up phase \citep{geppert1994,urpin1995,konar1997,Urpin1998OnPulsars}. The newly accreted material on the NS crust pushes the current-carrying layers towards deeper regions of higher density with higher conductivity. This slows down the field decay. The field ultimately freezes at a minimum strength $B_{\rm min}$ in the range of $10^7$G $\lesssim B_{\rm min} \lesssim 10^9$G. The value of $B_{\rm min}$ is correlated with the long-term accretion rate \citep{konar1997,Konar1999MagneticII,Konar1999MagneticIII,Halder2023DefiningPulsars}. We take $B_{\rm min} = 2 \times 10^8~$G for our numerical computations and show the effect of different $B$ values in Section \ref{subsec: Model parameters}. 

 Using the $\Mdot_2$ value obtained from the \mesa~calculations, we calculate $\Mdotin = \epsilon~ \Mdot_2$. Each time interval $dt$ and value of $M_1$ are calculated considering the current $\Mdotin$ value obtained with \mesa. Adjustable time-steps used by \mesa~could be too large ($dt > 10^2~$yr) to account for small $P$ changes of the NS. We use much shorter $dt$ values that can resolve the $P$ variations of the NS, which also affect the $\rin$ and torque calculations. Below we will describe the rotational evolution of NSs along the BD, UCXB, and MS tracks using illustrative model sources (Fig. \ref{fig:BDtrack_evl} – \ref{fig:MStrack_evl}).

\subsection{Rotational evolution of the NS on the BD-track}
\label{subsec:BDtrack_evl}

We obtain the model track seen in Fig. \ref{fig:BDtrack_evl} with initial period, $P_{\rm i} = 100~$s and $\Porbi = 1.5~$d. The evolution of the NS starts in the SU phase at $t \simeq 3.5~$Gyr with $\Mdotin \simeq 1 \x 10^{16}~$\gpers. The inner disc reaches the NS surface within $\simeq 1.2 \x 10^{-3}~$Gyr with $\Mdotin \simeq 3.5 \x 10^{17}~$\gpers~and $P \sim  0.1~$s. This situation persists for $\simeq 9.1 \x 10^{-2}~$Gyr effectively spinning the NS to $P \sim 2~$ms by $\Gamma_{\rm acc}$. The inner disc detaches from the NS surface when $P \simeq 1.9~$ms. A zoom-in of the detachment epoch is shown in Fig. \ref{fig:BDtrack_evl}c. The system enters the WP phase at $t \simeq 3.6~$Gyr. $\Mdotin$ remains close to its maximum value during the evolution, providing a strong $\Gamma_{\rm acc}$ spinning up the NS to its minimum $P \sim 1~$ms during the early stages of the WP phase. Afterwards, $\Mdotin$ decreases and the NS begins to spin down at $t \simeq 3.67~$Gyr below $\Mdotin \simeq 1 \x 10^{17}~$\gpers.

The RLOF stops ($\Mdotin = 0$) at $t \simeq 3.73~$Gyr and the system could be observable as an RMSP spinning down until $P \simeq 2.1~$ms (shaded area in Fig. \ref{fig:BDtrack_evl}). The RLOF resumes at $t = 4.6~$Gyr with $\Mdotin \sim 3.0 \x 10^{15}~$\gpers. The rotational evolution of the NS continues with the disc in the WP phase ($\rin = \rco$) slowing down mainly by $\Gamma_{\rm D}$ for the rest of its evolution. The system is a transient X-ray source during this late phase of evolution. $\Mdotin$ remains close to $\sim10^{15}~$\gpers~for another $\sim1~$Gyr. We observe small fluctuations (torque reversals) in $\Pdot$ which decrease while $\Mdotin$ remains at the same level and $P$ continues to increase from $P \sim 3~$ms to $P > 10~$ms during the long-term evolution (see Fig. \ref{fig:BDtrack_evl}b,d). Overall, $\Pdot$  decreases to a minimum value around $5 \x 10^{-20}~$\spers~and increases afterwards with decreasing $\Mdotin$. It is highly likely that sources evolving along the BD-track are observed as AMXPs during this evolutionary phase. For this illustrative source, at $t \sim 4.6~$Gyr, the properties of the binary and its components are in good agreement with the observed properties of AMXPs with BD donors. For a NS with $B = 2 \times 10^8~$G, which is the median value of RMSPs, and $\Porbi = 1.5~$d, our results indicate that LMXBs with BD companions can form within $8~$Gyr. Considering that donors in systems with smaller $\Porbi~$ have shorter lifetimes, we estimate  that some systems could form on a shorter time-scale.

\subsection{Rotational evolution of the NS on the UCXB-track}
\label{subsec:UCXBtrack_evl}

We obtain the illustrative long-term evolution of the NS along the UCXB-track with $P_{\rm i} = 100~$s and $\Porbi = 2.42~$d. For this model source (Case 1), the evolution has two main phases: an LMXB phase and a UCXB phase (See Fig. \ref{fig:UCXBtrack_evl}). The RLOF begins in the SU phase at $t \simeq 7.63~$Gyr with $\Mdotin\sim 2 \x10^{16}~$\gpers, and the inner disc moves inwards and reaches the NS surface with $\Mdotin \sim 1 \x10^{17}~$\gpers~within $\simeq 4.8 \x 10^{-3}~$Gyr. Starting from $t \simeq 7.64~$Gyr the NS rapidly spins up to $P \sim 1.7~$ms before the inner disc detaches from the NS surface at $t \simeq 7.68~$Gyr with $\Mdotin\sim 4.5 \x10^{17}~$\gpers. The SU phase continues for $\sim 1 \x 10^{-2}~$Gyr (see the zoom-in Fig. \ref{fig:UCXBtrack_evl}c). The WP phase begins when $\Mdotin$ decreases below $\sim 1 \x10^{17}~$\gpers. The NS spins up in the WP phase reaching its minimum $P \simeq 1.0~$ms around $t \simeq 7.72~$Gyr. As $\Mdotin$ decreases below $5 \x10^{16}~$\gpers, the system enters the transient regime and the NS starts to spin down while the system is still in the WP phase which persists for $\sim 2~$Gyr with $\Pdot \sim 2 \x10^{-19}~$\spers. The WP/SP transition takes place when  $\Mdotin \lesssim 1 \x 10^{11}~$\gpers~and $P \sim 12.4~$ms at $t \simeq 9.3~$Gyr. This marks the end of the LMXB phase, switching on the RMSP phase.

The UCXB phase starts with the onset of the second RLOF  at $t \sim 11.3~$Gyr. The $\Mdotin$ increases sharply from $\sim 1 \x10^{11}~$\gpers~to $\sim 1\x 10^{15}~$\gpers~within $\sim 1~$Myr taking the system into the SU phase. After the WP/SU transition, the inner disc arrives at the NS surface within $\sim  3 \x 10^{-3}~$Gyr ($\rin = r_{\star}$). Afterwards, the system accretes persistently while $\Mdotin$ increases sharply to near the Eddington limit, with $\Mdotin \sim 2 \x 10^{18}~$\gpers~on a time-scale of $\sim 1 \x 10^{-4}~$Gyr. During this epoch, the NS spins up rapidly from $P \sim 11.9~$ms to $P \sim 4.6~$ms. Subsequently, the inner disc detaches from the NS surface, and the system enters the WP phase when $\Mdotin \lesssim 2 \x 10^{16}~$\gpers. The minimum $P$ ($4.23~$ms) for the UCXB phase is reached during this phase at $t \simeq 11.4~$Gyr. The NS spins down in the WP phase reaching gradually to the properties of $P \sim 22~$ms, $\Pdot \sim 2 \x10^{-19}~$\spers, and $\Mdotin \sim 2.8 \x 10^{13}~$\gpers~at $t \simeq 14~$Gyr.

\begin{figure*}
    \centering
    \includegraphics[width=0.85\linewidth]{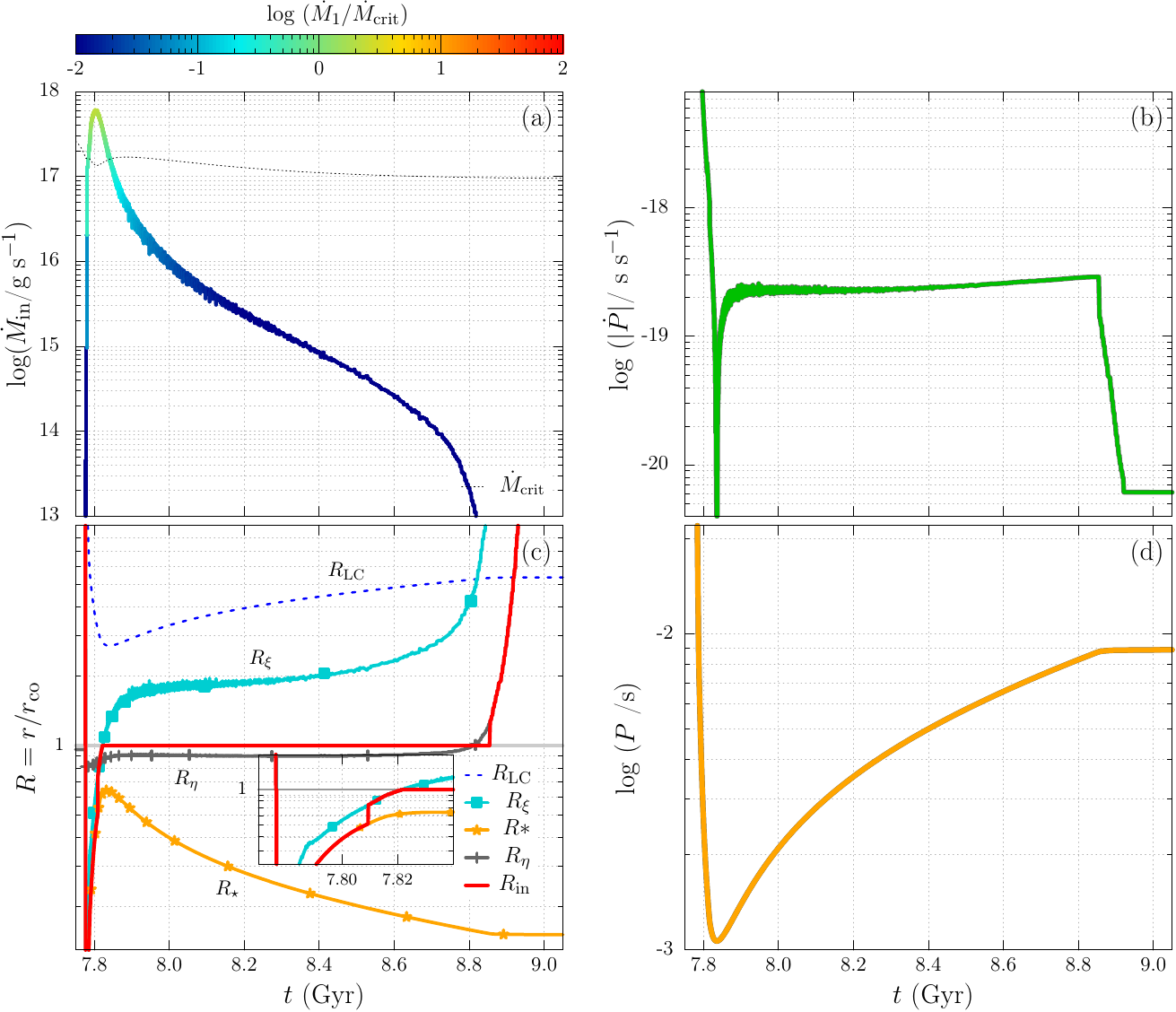}
    \caption{ Same as Fig. \ref{fig:BDtrack_evl} for a NS evolving on an MS-track with $\Porbi = 2.6~$d.}
    \label{fig:MStrack_evl}
\end{figure*}

Fig. \ref{fig:UCXBtrack_evl}c,d illustrates evolutions with different $\Porbi$ values corresponding to the three different cases discussed in Section \ref{subsec:UCXBtrack}. For Case 1 ($\Porbi = 2.42~$d), the NS is in the WP phase with  $\Mdotin \sim 5 \x 10^{14}~$\gpers, $P \sim 6.0~$ms,~and $\Pdot \sim 2 \x10^{-19}~$\spers~between $t \simeq 11.5~$Gyr and $t \simeq 12.0~$Gyr. The observable properties of the system (binary, NS, and donor) at this epoch estimated by the model are very similar to observed properties of AMXPs with He WD donors. Similar observable properties are simultaneously observed between $t \simeq 8.3~$Gyr and $t \simeq 9.0~$Gyr for Case 2 ($\Porbi = 2.20~$d), and between $t \simeq 7.0~$Gyr and $t \simeq 7.8~$Gyr for Case 3 ($\Porbi = 1.90~$d) (see Fig. \ref{fig:UCXBtrack_evl}b,d). Our results indicate that AMXPs that evolve on the UCXB-track can form within $\sim 8 ~-~12~$Gyr.

\begin{figure*}
    \centering
    \includegraphics[width=0.9\linewidth]{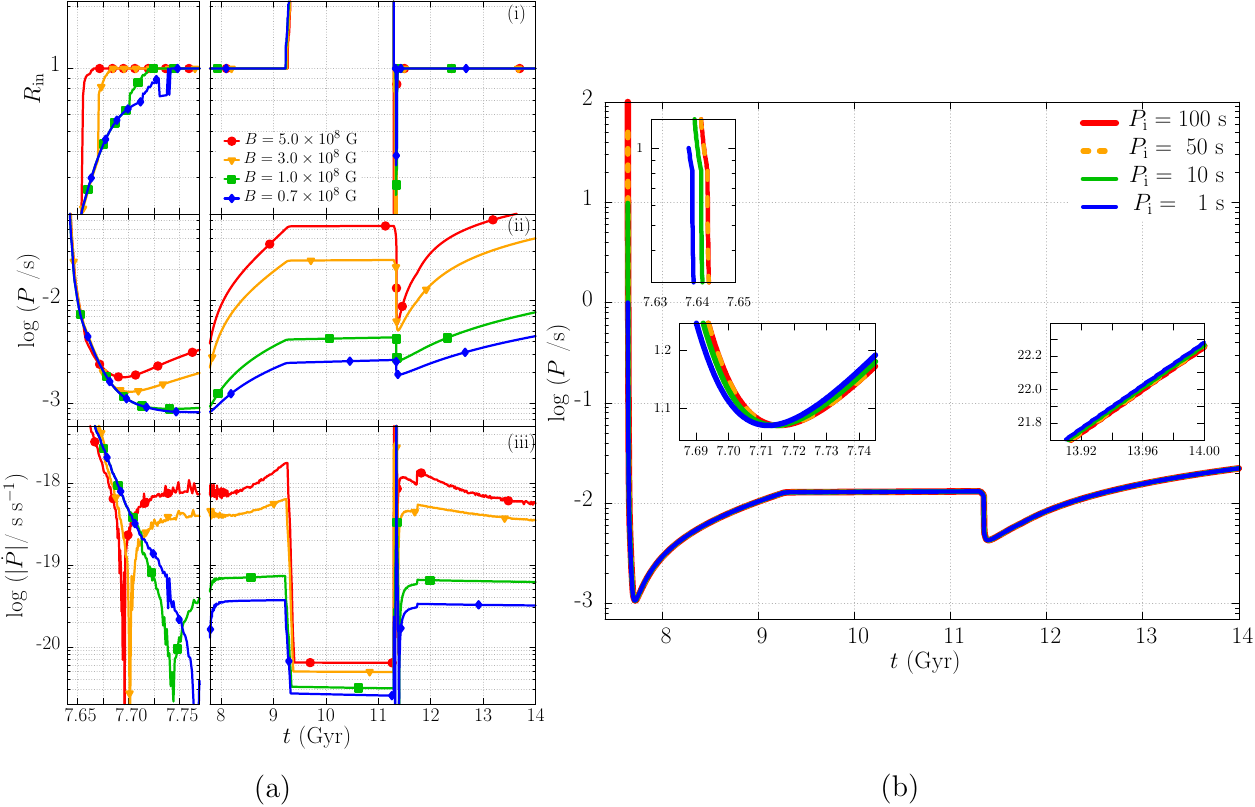}
    \caption{(a) Effect of different $B$ values on the long-term evolution of $\Rin$, $P$, and $\Pdot$ for $P_{\rm i} = 100~$s. (b) Effect of $P_{\rm i}$ on the long-term evolution of $P$ for $B = 2 \x 10^{8}~$G. The model curves in (a) and (b) are obtained using the $\Mdotin$ history of the binary model with $\Porbi = 2.42~$d (Case 1 in Section \ref{subsec:UCXBtrack_evl}), $\xi = 0.8$, $\eta = 0.8$, and $\DeltaR = 0.2$.}
    \label{fig:model_parameters}
\end{figure*}

\subsection{Rotational evolution of the NS on the MS-track}
\label{subsec:MStrack_evl}

The illustrative model curves seen in Fig. \ref{fig:MStrack_evl} are obtained with $P_{\rm i} = 100~$s and $\Porbi = 2.6~$d. The rotational evolution of this model source has the same morphology with the LMXB phase of the UCXB-track with $\Porbi = 2.42~$d (see Case 1, Section \ref{subsec:UCXBtrack_evl}). RLOF (in the SU phase) begins at $t \simeq 7.77~$ Gyr with $\Mdotin\sim 2 \x10^{16}~$\gpers. The inner disc penetrates into $\rco$ and reaches the surface of the NS within about $3 \x 10^{-3}~$Gyr. While $\rin = r_{\star}$ and $\Mdotin$ increases to $\sim 5\x10^{17}~$\gpers, the NS spins up to $P \sim 1.56~$ms until the inner disc detaches from the NS surface at $t \simeq 7.80~$ Gyr. The spin-up continues until after the SU/WP transition (see the zoom-in Fig. \ref{fig:MStrack_evl}c) and $P = 1.0~$ms and $\Mdotin  \sim 1 \x10^{17}~$\gpers~at $t \simeq 7.83~$ Gyr. For this illustrative source, the minimum $P$ is reached during the WP phase. Afterwards, the NS spins down with further decrease in $\Mdotin$, and the system enters the transient regime. During this epoch, $\Gamma_{\rm D}$ becomes the dominant torque with $\Pdot \sim 2 \x10^{-19}~$\spers~ (see Fig. \ref{fig:MStrack_evl}d). The WP phase persists for $\sim 1~$ Gyr across a large range of $\Mdotin$. The NS enters the SP phase when $\Mdotin$ falls below $\sim 10^{11}~$\gpers~at $t \simeq 8.8~$ Gyr. Eventually, the RLOF ceases. The NS could be observed as an RMSP after the WP/SP transition. It is seen in Fig. \ref{fig:MStrack_evl} that the source has $P \sim 2.64~$ms, $\Pdot \sim 2 \x10^{-19}~$\spers, and $\Mdotin\sim 4.5 \x10^{15}~$\gpers~at $t \simeq 8.1~$ Gyr while the system is evolving in the WP phase as a transient source. The state of the NS and the binary properties at this epoch are in good agreement with the observed properties of AMXPs with MS donors, which suggests that the model MS-track with $\Porbi = 2.6~$d is a reasonable evolutionary path for AMXPs.

\subsection{Effect of $P_{\rm i}$ and $B$ on the NS evolution}
\label{subsec: Model parameters}

The model curves seen in Figs \ref{fig:BDtrack_evl} – \ref{fig:model_parameters} are produced with the same parameters ($\xi = 0.8$, $\eta = 0.8$, and $\DeltaR = 0.2$). A detailed study for these parameters can be found in \cite{Ertan2021}. We use the illustrative model binary, Case 1 of the UCXB-track (see Section~\ref{subsec:UCXBtrack_evl}) to show the effect of $B$ on $\Rin$, $P$, and $\Pdot$ in Fig.~\ref{fig:model_parameters}(a), and the effect of $P_{\rm i}$ on the evolution in Fig.~\ref{fig:model_parameters}(b). It is seen in Fig.~\ref{fig:model_parameters}(a) that higher (lower) $B$ values result in the earlier (later) SU/WP transitions occurring at higher (lower) $\Mdotin$ levels. Additionally, NSs with stronger dipole fields evolve towards longer periods due to the $B^2$ dependence of the spin-down torques, $\Gamma_{\rm D}$ and $\Gamma_{\rm dip}$.

The long-term rotational evolution of model NSs shown in Figs \ref{fig:BDtrack_evl}–\ref{fig:MStrack_evl} is obtained with $P_{\rm i} = 100~$s. To see the effect of $P_{\rm i}$ on our results, we repeated the calculations using $P_{\rm i}=50~$s, $10~$s, and $1~$s for each of the cases presented in Section~\ref{subsec:UCXBtrack_evl}. Fig.~\ref{fig:model_parameters}(b) shows zoomed-in views of the spin evolution at three evolutionary stages. The top-left panel corresponds to the onset of RLOF, marking the beginning of the SU phase ($t\simeq 7.64~$Gyr). The centre-left panel shows the phase when $\Mdotin$ reaches its maximum ($t \simeq 7.71~$Gyr, $P \simeq 1.07~$ms). The centre-right panel corresponds to the end of the evolution ($t \simeq 14.00~$Gyr, $P \simeq 22.3~$ms). For the illustrative cases with $P_{\rm i} = 50~$s, $10~$s, and $1~$s, the resulting spin evolutions remain within $1\%$ of the reference model with $P_{\rm i} = 100~$s at all evolutionary stages seen in Fig.~\ref{fig:model_parameters}(b). We find that, like $P$, $\Pdot$ and $\Rin$ are also insensitive to $P_{\rm i}$.

\begin{figure}
    \centering
    \includegraphics[width=1.\columnwidth]{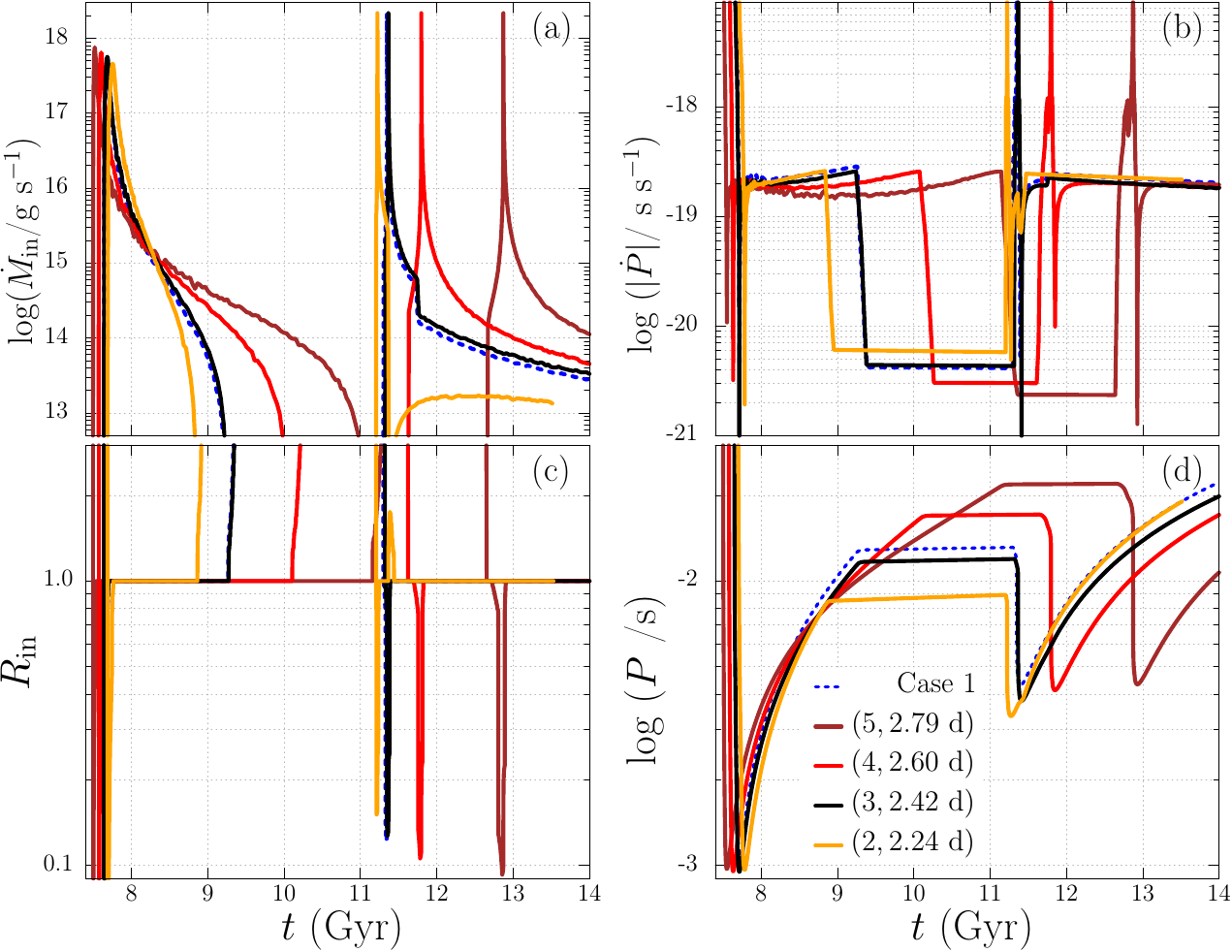}
    \caption{Same as Fig. \ref{fig:UCXBtrack_evl} for ($\gamamb, \Porbi$) pairs described in Section \ref{subsec:beta-gamma_binary} (also given in the figure) and $\beta = 0.4$ together with the evolution of Case 1 shown by the blue dashed curves.}
    \label{fig:Grid}
\end{figure}

\begin{figure*}
    \centering
    \includegraphics[width=1.0\linewidth]{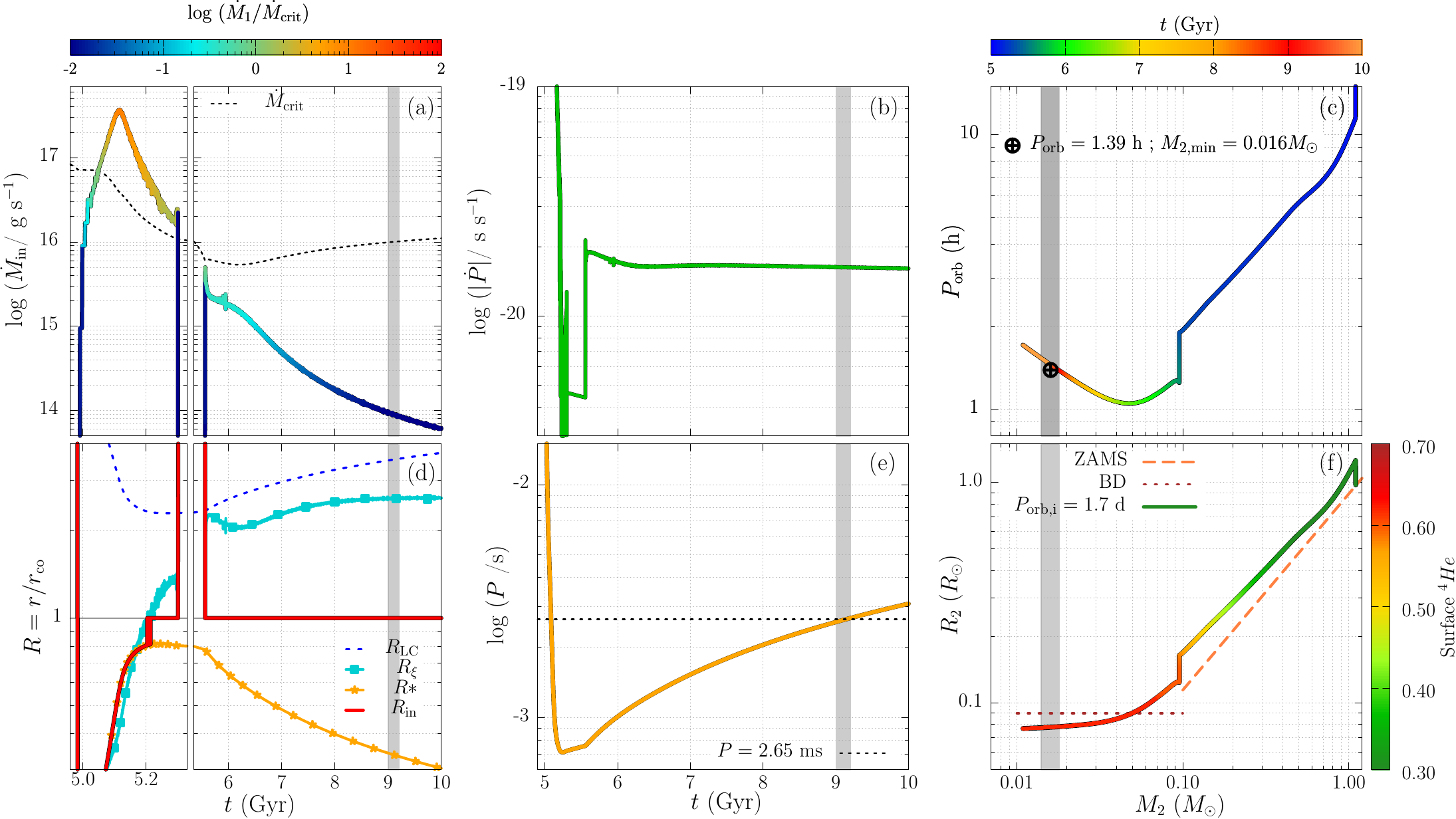}
    \caption{ Evolution of the binary and rotational properties of J1900 obtained with $\DeltaR = 0.2$, $\eta = 0.8$, $\xi = 0.8$, $B = 5 \times 10^7$~G, $P_{\rm i} = 100~$s and $\Porbi = 1.7~$d. (a) Evolution of $\Mdotin$ and $\Mdotcrit$ (dashed curve). The colour bar indicates $\dot{M_1}/ \dot{M}_{\mathrm{crit}}$. (b) Evolution of $\Pdot$. (c) Evolution of $\Porb$ as a function of $M_2$. The colour bar shows the time in Gyr. (d) Evolution of $\Rin$ (solid curve),  $R_{\rm LC}$ (dashed curve), $\Rxi$ (solid curve with squares), and $R_{\star}$ (solid curve with stars). (e) Evolution of the NS's $P$. (f) Evolution of $R_2$ as a function of $M_2$. The colour bar indicates the evolution of the helium surface abundance for the companion. The vertical grey strips seen in each panel indicate the time at which the binary and the NS properties of J1900 are obtained simultaneously.} 
    \label{fig:HETE}
\end{figure*}

\subsection{Effect of $\beta$ and $\gamamb$ on the NS evolution}
\label{subsec:beta-gamma_rotation}

Using the $\Mdot_2$ histories from the grid analysis of Section \ref{subsec:beta-gamma_binary}, we examine the effects of $\beta$ and $\gamamb$ on the rotational evolution of the NS. We again use Case 1 for illustration of these effects, as it passes twice through the SU, WP, and SP phases with a detachment phase in between. Fig.~\ref{fig:Grid} shows the long-term evolution of $\Mdotin$, $\Pdot$, $\Rin$, and $P$ for $\beta = 0.4$ and the same four ($\gamamb, \Porbi$) pairs used in Section \ref{subsec:beta-gamma_binary}. We also plot the evolution of Case~1 as our reference model for comparison. For each ($\gamamb, \Porbi$) pair, the variations of $\Rin$ with $\Mdotin$ (Fig.~\ref{fig:Grid}a,c) in the SU, WP, and SP phases are consistent with those described in Section~\ref{subsec:UCXBtrack_evl}. Similarly, variations in $\Pdot$ remain within a factor of two of the values obtained in Case 1 (Fig.~\ref{fig:Grid}b). $\Rin$ and $\Pdot$ are moderately sensitive to $\beta$ and $\gamamb$ values, while the $P$ evolution is highly sensitive to these parameters (Fig.~\ref{fig:Grid}d). During the first spin-up episode, $P$ reaches a minimum of $\sim 1~$ms for all $\gamamb$. The values obtained for $\beta = 0.2, 0.6$, and $0.8$ remain within $3 \%$ of the minimum values obtained for $\beta = 0.4$. During intermediate stages of the evolution (pre-detachment), we find that models with higher (lower) $\gamamb$ values experience a stronger (weaker) spin-down as they spend more (less) time in the WP phase. Following the detachment phase, the system enters a second SU phase where the minimum period attained is $P \sim 4~$ms with $\beta$-dependent deviations rising to $12-22 \%$. This trend reverses at the end of the evolution, where longer final periods ($P \sim 20~$ms) are attained with lower $\gamamb$, and the total dispersion across the $\beta$ range reaches approximately $40 \%$. Overall, the varying effects of $\gamamb$ and $\beta$ on the rotational evolution are a direct consequence of how the variations of these parameters affect $\Mdotin$ as we have shown in Section \ref{subsec:beta-gamma_binary}.

\section{Applications to the observed AMXPs}
\label{Section4}

\begin{figure*}
    \centering
    \includegraphics[width=1.\linewidth]{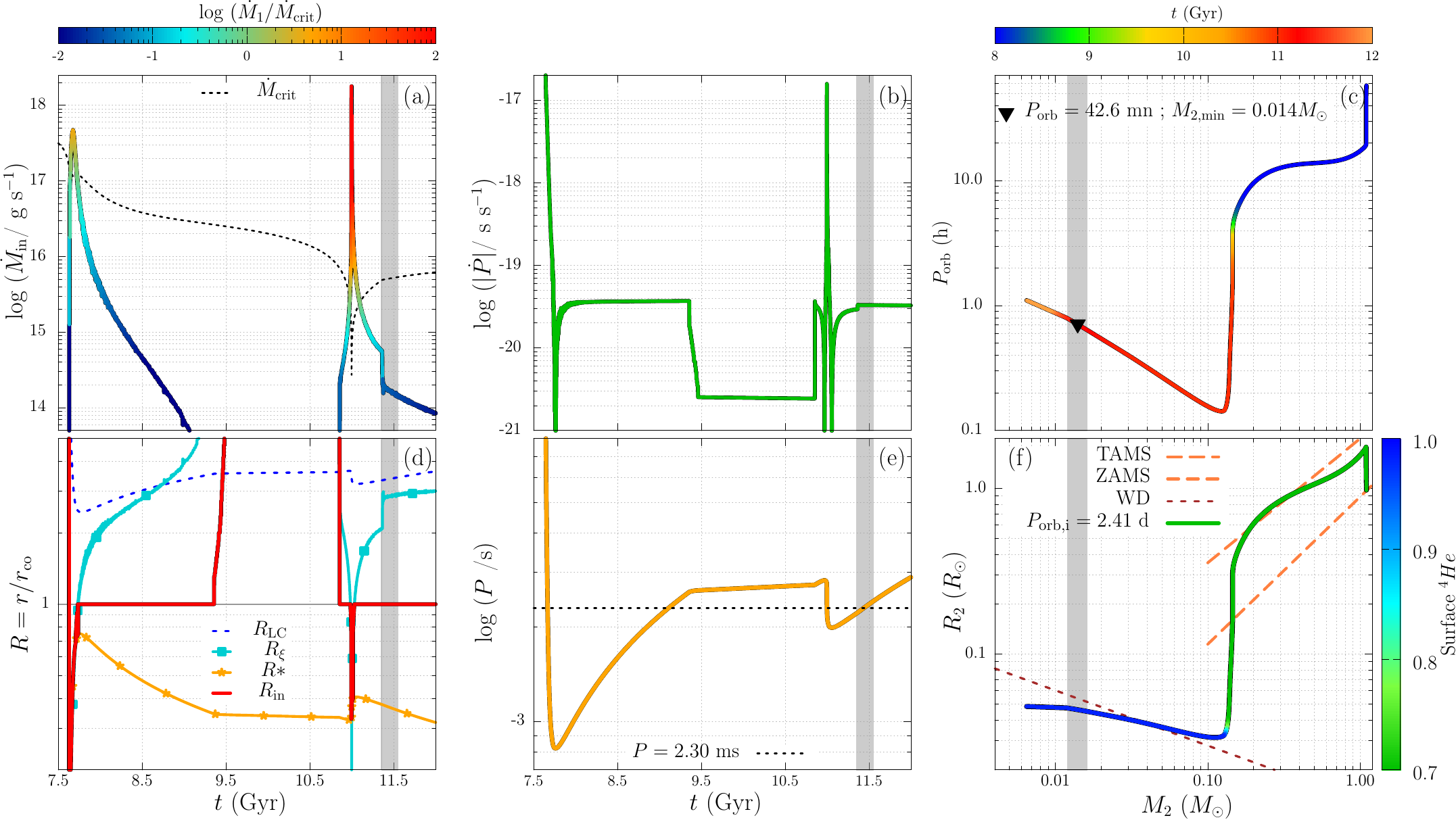}
    \caption{Evolution of the binary and rotational properties of J1751. Same as Fig. \ref{fig:HETE}, except $\Porbi = 2.41$~d, and $B = 7 \times 10^7$~G. The vertical grey strips seen in each panel indicate the time at which the binary and the NS properties of J1751 are obtained simultaneously.}
    \label{fig:J1751}
\end{figure*}

Here, we will present the applications of our analytical model together with the \mesa~code to the binary and NS properties of the three AMXPs selected from each of the three main binary tracks discussed in Section \ref{Section3}. The model parameters are given in figure captions (Figs. \ref{fig:HETE}–\ref{fig:Aql}). Using the observed properties of the sources that are available in the literature, we calculate the relative errors ($\theta (P) = |P- P_{\rm obs}| / P_{\rm obs}$) for each of these values calculated in our model. We look for a solution set of $t, P, \Pdot, \Porb, M_2 $, and $\Mdotin$ values for which the four criteria mentioned in Section \ref{subsec:The evolutionary tracks of AMXPs} are met and $\theta (P)$ is equal to its minimum. Afterwards, we discuss the relative errors $\theta (\Pdot)$, $\theta (\Porb)$, $\theta (M_2)$, and $\theta (\Mdotin)$ in other observables. The evolution of the NS and the binary properties are given in Figs \ref{fig:HETE}–\ref{fig:Aql}. In these figures, vertical grey strips show the time interval during which the model reproduces the observed NS and binary properties simultaneously. We select $B$ values in the range of $10^7$ to $10^9$~G and take $\xi = 0.8$, $\eta = 0.8$, and $\DeltaR = 0.2$ to study the long-term evolution of HETE J1900.1-2455 (hereafter, J1900) for the BD-track, XTE J1751-305 (hereafter, J1751) for the UCXB-track, and Aql X-1 for the MS-track. Our results are summarized in Table \ref{tab:AMXPresults_final}.

\subsection*{HETE J1900.1-2455}
\label{disc:J1900}

The source J1900 was discovered during its X-ray outburst in June 2005 \citep{Vanderspek2005PossibleHETE}, and coherent $2.65$ ms X-ray pulsations were subsequently detected in July 2005 \citep{Kaaret2006DiscoveryJ1900.12455}. The donor is a low-mass BD with $M_{\rm 2,min} = 0.016 \Msun~$and $\Porb = 1.39~$h. J1900 exhibits sporadic X-ray pulsation behaviour, classifying it as an intermittent AMXP \citep{DiSalvo2022AccretionPulsars}. The distance to the source, $d = 4.7 \pm 0.6$~kpc, is estimated from the photospheric radius expansion (PRE) during an X-ray burst \citep{Galloway2020}. The source remained in outburst for a decade, until October 2015 \citep{Heinke2025CatalogBinaries}.

We obtain reasonable model curves that could represent the evolution of J1900, taking $M_{\mathrm{1,i}} = 1.4 ~\Msun$, $M_{\mathrm{2,i}} = 1.1 ~\Msun$, $\gamamb = 3$, $\beta = 0.5$, and $\Porbi = 1.7$~d. For the NS evolving with the corresponding $\Mdotin$ history, we take $P_{\rm i} = 100$~s and $B = 5 \times 10^7$~G. At $t \simeq 9.1~$Gyr, we obtain a solution for which $\theta (P) = 4.29 \x 10^{-6}$ (indicated by the grey strip in Fig. \ref{fig:HETE}), $\Porb \simeq 1.41~$h ($\theta (\Porb) = 0.013$), and the donor is a hydrogen-poor BD with $M_{\rm 2} \simeq 0.016 ~\Msun$ ($\theta (M_2) = 0.081$). At this stage, the NS is in the WP phase with $M_1 = 1.93~\Msun$, $\Mdotin \simeq 8.6 \times 10^{13}~$\gpers, and $\Pdot \simeq 1.6 \times 10^{-20}~$\spers~which is also in agreement with the long-term secular $\Pdot$ values observed in AMXPs during their quiescent states \citep{DiSalvo2022AccretionPulsars, Patruno2021AccretingPulsars}.

\begin{table}
    \centering
    
    \caption{Comparison of the observed properties ($P, \Pdot, \Porb, M_2$, and $\Mdotin$) with the model sources for three AMXPs. The $\theta$ values show the relative errors described in Section \ref{Section4}.}
    \label{tab:AMXPresults_final}
    \resizebox{1.\columnwidth}{!}{
    \begin{tabular}{l|c|c|c|c}
        \hline
        \textbf{Property}   & \textbf{}         & \textbf{J1900}        & \textbf{J1751} & \textbf{Aql X-1} \\
        \hline                                                          
                            & Observed          & $2.6504377 $          & $2.2971710$       & $1.817289$ \\
       $P$ (ms)             & Model             & $2.6504490$           & $2.2968364$         & $1.817543$ \\
                            & $\theta(P)$       & $4.29 \x 10^{-6}$     & $1.46 \x 10^{-4}$ & $1.39 \x 10^{-4}$ \\
        \hline
                            & Observed          & 1.39                  & 0.71             & 18.95 \\
      $\Porb$ (h)           & Model             & 1.41                  & 0.81             & 19.90 \\
                            & $\theta(\Porb)$   & 0.013                 & $0.15$           & 0.05 \\
        \hline
                            & Observed          & 0.016                 & 0.014             & 0.56 \\
      $M_2$ ($~\Msun$)      & Model             & 0.017                 & 0.011             & 0.38 \\
                            & $\theta (M_2) $   & 0.081                 & 0.184             & 0.32 \\
        \hline
                            & Observed          & $-$                   & $3.21_{-1.83}^{+1.64} \times 10^{14}$ & $1.04 \times 10^{17}$ \\
      $\Mdotin$ (\gpers)   & Model             & $ 8.6\times10^{13}$ & $1.73\times 10^{14}$ & $1.22 \times 10^{17}$ \\
                            & $\theta(\Mdotin)$ & $-$                   & 0.459             & 0.08 \\
        \hline
                            & Observed          & $-$                   & $2.9 \x 10^{-20}$ & $-$ \\
     $\Pdot$ (\spers)       & Model             & $1.6 \times 10^{-20}$ & $3.2 \x 10^{-20}$ & $6.1 \times 10^{-19}$ \\
                            & $\theta (\Pdot)$  & $-$                   & $0.136$ & $-$ \\
        \hline
        $t_{\rm solution}$ (Gyr) &  & 9.14 & 11.43 & 7.92 \\
        \hline
    \end{tabular}
                        }
\end{table}

\subsection*{XTE J1751-305}
\label{disc:J1751}

The source J1751 was discovered in 2002 with the detection of $P = 2.30$~ms X-ray pulsations during an X-ray outburst \citep{Markwardt2002DiscoveryJ1751305}. The source showed outbursts again in 2005, 2007, and 2009, without any X-ray burst. The distance to the source is estimated to be $ \sim 8.5$~kpc \citep{Heinke2025CatalogBinaries}. $\Porb = 42.3~$min and the donor is a low-mass He WD with $M_{\rm 2,min} = 0.013 - 0.017~ \Msun$ \citep{Deloye2003WhiteObjects}. We model the long-term rotational evolution of J1751 with $M_{\mathrm{1,i}} = 1.4 ~\Msun$, $M_{\mathrm{2,i}} = 1.1 ~\Msun$, $\gamamb = 3$, $\beta = 0.5$, and $\Porbi = 2.41$~d. For the NS we take $P_{\rm i} = 100$~s and $B = 7 \times 10^7$~G. We find a reasonable solution at $t \simeq 11.4~$Gyr with $\theta (P) = 1.46 \x 10^{-4}$ (indicated by the grey strip in Fig. \ref{fig:J1751}), when the NS has $P = 2.30~$ms, $\Porb \simeq 48.8~$min ($\theta (\Porb) = 0.15$), and the donor is a He WD with $M_{\rm 2} \simeq 0.014 ~\Msun$ ($\theta (M_2) = 0.184$). The NS is in the WP phase, with $M_1 = 1.93 ~\Msun$, $\Mdotin \simeq 1.73 \times 10^{14}~$\gpers~ ($\theta (\Mdotin) = 0.459$), and $\Pdot~\simeq~3.29 \times 10^{-20}~$\spers~($\theta (\Pdot) = 0.136$). At this stage of the evolution, time-averaged $\Mdotin =3.21_{-1.83}^{+1.64} \times 10^{14}~$\gpers~ estimated from about ten-years long observations \citep{Heinke2013GalacticEvolution.}. The measurements between the 2002 and 2009 outbursts give $\Pdot = (2.9 \pm{0.6}) \times 10^{-20}$~\spers~ \citep{Riggio2011SecularJ1751-305}. It is seen in Fig. \ref{fig:J1751} that both measurements are in good agreement with the properties of the model at $t \simeq 11.4~$Gyr. In Fig. \ref{fig:J1751}e, the $P$ curve estimated in the model is given by the solid curve and the dashed curve indicates the $P$ of J1751. The two curves intersect four times during the rotational evolution of the source, but only the last intersection simultaneously matches the current NS and binary properties of J1751 in the WP phase; this is not the case for the first three intersections.

\subsection*{Aql X-1}
\label{disc:Aql}

\begin{figure*}
    \centering
    \includegraphics[width=\linewidth]{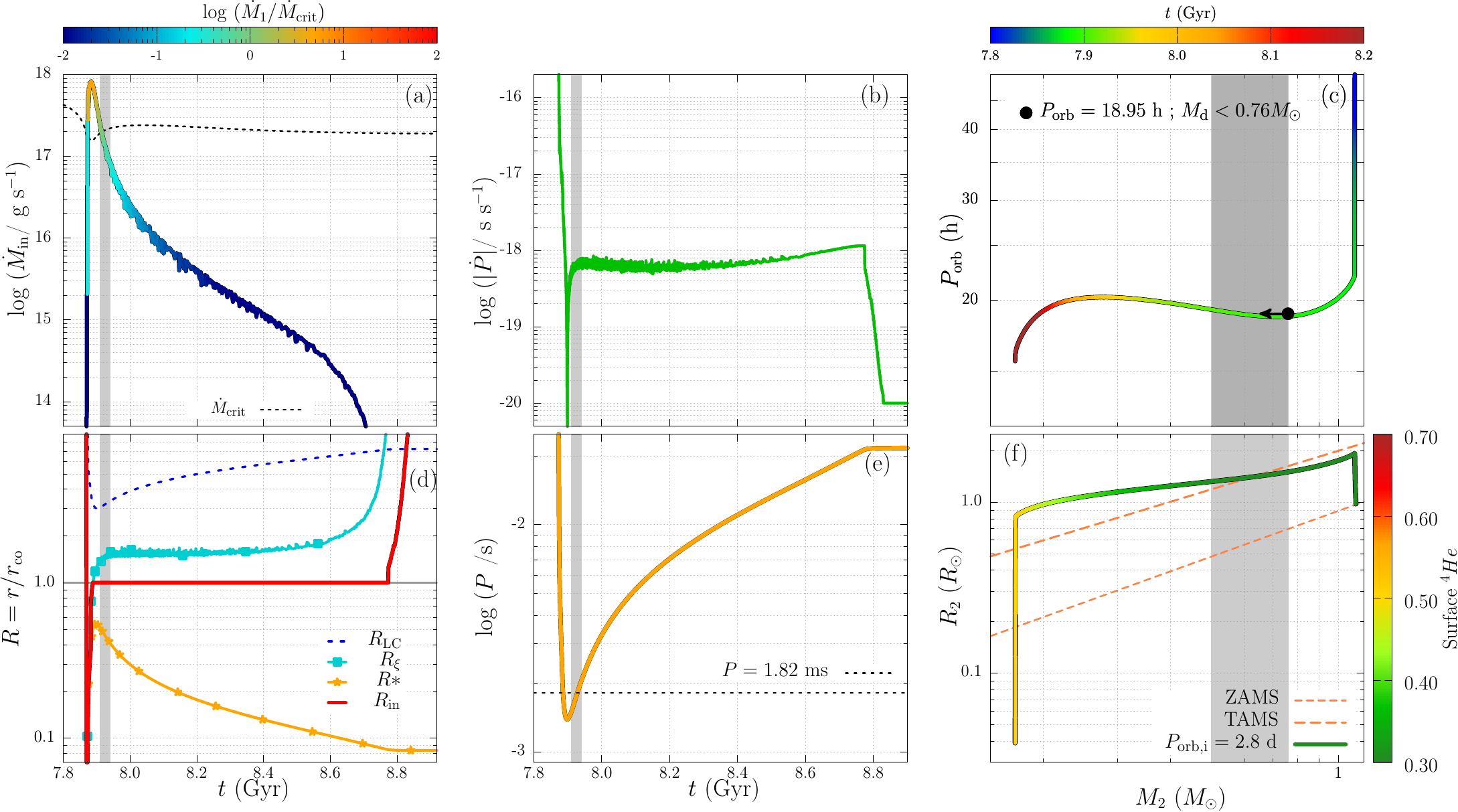}
    \caption{Evolution of the binary and rotational properties of Aql X-1. Same as Fig. \ref{fig:HETE}, except $\Porbi = 2.8$~d, and $B = 4 \times 10^8$~G. The vertical grey strips seen in each panel indicate the time at which the binary and the NS properties of Aql X-1 are obtained simultaneously.}
    \label{fig:Aql}
\end{figure*}

Aql X-1 was discovered in 1991 with $\Porb = 18.95~$h \citep{Chevalier1991DiscoveryX-1.} and a $K4 \pm 2~$MS donor star with $M_{2} < 0.76 ~\Msun$ \citep{Sanchez2017TheSpectroscopy} during an X-ray outburst. Aql X-1 is observed in annual outbursts with occasional X-ray bursts. The distance is estimated to be $\sim 4.5~$kpc from the PRE \citep{Galloway2008ThermonuclearExplorer}. Coherent $P = 1.82$~ms X-ray pulsations that lasted for $\sim 150~$s were recovered from archival data making the source an intermittent AMXP \citep{Casella2008DiscoveryX-1}. The model curves seen in Fig. \ref{fig:Aql} are obtained with $M_{\mathrm{1,i}} = 1.4 ~\Msun$, $M_{\mathrm{2,i}} = 1.1 ~\Msun$, $\gamamb = 3$, $\beta = 0.5$, and $\Porbi = 2.80$~d. Using the corresponding $\Mdotin$ history, we obtain the NS evolution seen in Fig. \ref{fig:Aql} with $P_{\rm i} = 100$~s and $B = 4 \times 10^8$~G. For this illustrative evolution, the NS acquires $P = 1.82~$ms ($\theta (P) = 1.39 \x 10^{-4}$), $\Porb = 19.9~$h ($\theta (\Porb) = 0.05$) while the donor is an MS star with $M_{\rm 2} \simeq 0.4~\Msun$ ($\theta (M_2) = 0.323$) at $t \simeq 7.9~$Gyr. All these binary and NS properties predicted by the model for $t \simeq 7.9~$Gyr are close to the values estimated from the observations. In the model, the MS value at $t \simeq 7.9~$Gyr is also close to the reported $M_{\rm 2} \simeq 0.6 ~\Msun$ for an inclination sin $(i) = 1$ \citep{DiSalvo2022AccretionPulsars}. At this time of evolution, we find the NS in the WP phase as a transient source with $M_1 = 1.75~\Msun $, $\Mdotin \simeq 1.2 \times 10^{17}~$\gpers~ ($\theta (\Mdotin) = 0.08$), and $\Pdot \simeq 6.1 \times 10^{-19}~$\spers.

Table \ref{tab:AMXPresults_final} compares the model results with the observed properties of the three sources. It is seen that the model yields reasonable results that are consistent with the binary and individual NS properties for the three AMXP sources representing the three main evolutionary tracks. The observed properties and long-term spin-down behaviour of these three sources can be consistently reproduced within the WP phase during their transient evolution. In an independent work, the same torque model was applied to five AMXPs with known secular $\Pdot$ values to investigate the torques acting on these systems throughout their outburst-quiescence cycles (Ertuğrul et al. 2026, in preparation). Their results are consistent with the observed properties of both the outburst and quiescent states, together with the resultant secular spin-down of these AMXPs. In future work, the $\Mdotin$ histories obtained from \mesa~will be used to investigate the rotational evolution of these sources, considering their transient $\Lx$ behaviours.

\section{Summary \& discussion}
\label{Section5}

Using reasonable \mesa~model parameters ($\beta$ and $\gamamb$) similar to those employed in earlier works \citep{Gossage2023MagneticRegimes,Kar2024Long-termPulsars}, we obtained binary evolutions leading to the properties of the three basic AMXP groups classified by their companions (BD, He WD, and MS). Our results are consistent with the observed binary properties of AMXPs. Next, we used these $\Mdotin$ histories obtained from the \mesa~models as inputs into our torque model to calculate the rotational evolution of NSs evolving along the BD, UCXB, and MS tracks. We have explained the typical evolutionary stages and properties of AMXPs evolving on these tracks, using illustrative sources (Figs \ref{fig:BDtrack}–\ref{fig:MStrack}). We also described the rotational evolutionary phases and the inner disc properties of the NSs accompanying the binary evolution (Figs \ref{fig:BDtrack_evl}–\ref{fig:MStrack_evl}). Next, we applied the model to three sources selected from the three groups of AMXPs. The binary properties ($\Porb$ and $M_2$) and the individual source properties ($P$ and $\Pdot$) are produced simultaneously for each of these sources (Figs \ref{fig:HETE}–\ref{fig:Aql}).

In our binary analysis, all simulated models are consistent with the observational constraints defining viable AMXP evolutionary tracks. These systems evolve into transient sources within a Hubble time and simultaneously reproduce the observed ranges of $M_2$, $\Porb$, surface composition, and donor type for each of the three groups of AMXPs. To assess the effect of key \mesa~parameters, we simulated $16$ binary models across a grid of $\beta$ and $(\gamamb,\Porbi)$ pairs. Our results show that while $\beta$ influences the efficiency of mass transfer, the overall evolutionary behaviour—especially the $\Mdot_2$ histories—is more sensitive to $\gamamb$. Nevertheless, we find that $\Porbi$ remains the most influential parameter in determining the evolutionary track, as variations in $\Porbi$ can lead to fundamentally different donor types and final system properties. Within this framework, our combined binary and NS evolution favours $\gamamb = 3$--$5$, whereas models with $\gamamb = 2$ often require $\Porbi$ tuning.

With reasonable values of the model parameters ($\xi, \eta$, and $\DeltaR$), we find that the rotational evolution of the NS is not sensitive to $P_{\rm i}$ while it is very sensitive to $B$. Extending the results of our grid analysis to the rotational evolution of the NS, we find that the impact of $\gamamb$ and $\beta$ on the evolution directly reflects their influence on $\Mdotin$ via $\Mdot_2$.

In this study, we employed the $\tau$-boosted MB prescription of VIH19 to obtain the $\Mdotin$ histories of AMXPs, whereas \citet{Kar2024Long-termPulsars} adopted the conventional MB prescription of \citet{Rappaport1983ABraking.}. To order of magnitude, the resulting $\Mdotin$ histories in the two studies are comparable. The most notable differences arise in the predicted NS rotational evolution. We argue that these differences are primarily due to $\rin$ and associated torque calculations. Using the torque model of \citet{Bhattacharyya2017THEPULSARS}, \citet{Kar2024Long-termPulsars} could not reproduce the rotational properties of fast-spinning AMXPs ($P \lesssim 3~$ms). Their study, which extensively explores the $(M_{\rm 2,i},\,\Porbi)$ parameter space for convergent LMXBs, successfully reproduces the binary properties of these systems while it fails to account for their NS rotational properties. They suggested that the transient accretion during the late evolutionary stages may solve this problem. By adopting the torque model of \citet{Ertan2021}, we were able to reproduce both the binary evolution and the NS rotational properties of the three AMXP groups self-consistently, by varying only $\Porbi$ for reasonable $B$ and other model parameters.

\section{Conclusions}
\label{Section6}

We have investigated the long-term binary and accompanying NS evolution in LMXBs, focusing on the formation of AMXPs along three distinct evolutionary tracks that depend on the donor type: BD-track (BD donor), UCXB-track (He WD donor), and MS-track (MS donor). For an evolutionary model track to be counted as a reasonable representation of the evolution into one of these LMXB families, the model should simultaneously reproduce the $M_2$, $R_2$, $\Porb$ values, and the surface composition consistent with the observations. Using the $\Mdotin$ evolution estimated from these binary evolutions, we have analysed the evolution of the NS on these tracks separately. Our results on the binary evolution obtained using the \mesa~code can be summarized as follows. (1) the $\tau$-boosted MB model yields evolutionary curves that are consistent with observed $\Lx$ levels of LMXBs, in agreement with the earlier studies \citep{Van2019Low-massPrescription, Deng2021EvolutionPrescriptions}. (2) Among the initial conditions, $\Porbi$ is the most influential factor affecting the binary evolution. (3) $\gamamb$ which affects the duration of a mass transfer epoch plays a greater role than $\beta$ which affects the efficiency of the mass transfer. (4) With slight adjustments to $\Porbi$, evolutionary paths with same $\Mdot_2$ morphology can be obtained  for different $\gamamb$ and $\beta$ values. (5) Our results favour $\gamamb \geq 3$ values since $\gamamb =2$ may require parameter tuning to obtain results consistent with observations. The results of our evolutionary tracks are in good agreement with the observed properties of AMXPs, except for three AMXPs with BD donors. We have estimated that the effect of irradiation of the companion by the NS, which was not taken into account in this work, could account for these sources as well. The irradiation effect on the evolution of these systems will be studied in an independent work. Using the $\Mdotin$ obtained with \mesa~for the three evolutionary tracks of AMXPs, we have employed the analytical model developed by \cite{Ertan2021} to calculate the $\rin$ and the torques acting on the NS during the long-term evolution. Our results show that the model can account for the evolution of the NSs in these systems in accordance with their binary evolution. Our model naturally accounts for the fast-spinning AMXPs ($P \lesssim 3~$ms) with degenerate donors, which \citet{Kar2024Long-termPulsars} found challenging to reproduce.  With reasonable parameters, the model reproduces the observed properties ($P$ and $\Pdot$) of HETE J1900.1-2455, XTE J1751-305, and Aql X-1 representing the three evolutionary tracks, simultaneously with the donor type, $M_2$, and $\Porb$. Using the \mesa~code, a similar study is required to investigate the formation of millisecond pulsars above the $\Pbif$ of AMXPs. Such evolutionary tracks that lead to the formation of divergent systems namely, RMSPs with $\Porb > 10~$d are a large group within the binary pulsar population.

\section*{Acknowledgements}

 We thank the anonymous referee, for very useful comments that have considerably improved our manuscript. NN thanks Yuki Kaneko for useful comments that have improved the manuscript. We acknowledge research support from Sabanc{\i} University, and from T\"{U}B\.{I}TAK (The Scientific and Technological Research Council of Turkey) through grant 123F083. 


\section*{Data Availability}

The {\tt MESA} inlists used in this study can be shared upon request.



\bibliographystyle{mnras}
\bibliography{references} 

@article{Wijnands1998ASystem,
    title = {{A millisecond pulsar in an X-ray binary system}},
    year = {1998},
    journal = {Nature 1998 394:6691},
    author = {Wijnands, Rudy and Van Der Klis, Michiel},
    number = {6691},
    pages = {344--346},
    volume = {394},
    publisher = {Nature Publishing Group},
    url = {https://www.nature.com/articles/28557},
    doi = {10.1038/28557},
    issn = {1476-4687}
}

@article{Alpar1982,
    title = {{A new class of radio pulsars}},
    year = {1982},
    journal = {Nature},
    author = {Alpar, M. A. and Cheng, A. F. and Ruderman, M. A. and Shaham, J.},
    number = {5894},
    pages = {728--730},
    volume = {300},
    publisher = {Nature Publishing Group},
    url = {https://www.nature.com/articles/300728a0},
    doi = {10.1038/300728a0},
    issn = {1476-4687}
}

@article{Rappaport1983ABraking.,
    title = {{A new technique for calculations of binary stellar evolution application to magnetic braking.}},
    year = {1983},
    journal = {ApJ},
    author = {Rappaport, S and Verbunt, F and Joss, P C and Rappaport, S and Verbunt, F and Joss, P C},
    pages = {713--731},
    volume = {275},
    publisher = {American Astronomical Society},
    url = {https://ui.adsabs.harvard.edu/abs/1983ApJ...275..713R/abstract},
    doi = {10.1086/161569},
    issn = {0004-637X}
}

@article{srinivasan1990,
    title = {{A novel mechanism for the decay of neutron star magnetic fields}},
    year = {1990},
    journal = {Current Science},
    author = {Srinivasan, G and Bhattacharya, D and Muslimov, A.~G. and Tsygan, A.~J.},
    pages = {31--38},
    volume = {59}
}

@article{Archibald2009ALink,
    title = {{A Radio Pulsar/X-ray Binary Link}},
    year = {2009},
    journal = {Science},
    author = {Archibald, Anne M and Stairs, Ingrid H and Ransom, Scott M and Kaspi, Victoria M and Kondratiev, Vladislav I and Lorimer, Duncan R and McLaughlin, Maura A and Boyles, Jason and Hessels, Jason W T and Lynch, Ryan and van Leeuwen, Joeri and Roberts, Mallory S E and Jenet, Frederick and Champion, David J and Rosen, Rachel and Barlow, Brad N and Dunlap, Bart H and Remillard, Ronald A},
    number = {5933},
    pages = {1411--1414},
    volume = {324},
    url = {https://www.science.org/doi/10.1126/science.1172740},
    doi = {10.1126/science.1172740},
    issn = {0036-8075},
    arxivId = {0905.3397}
}

@article{Bassa2014AXSSJ122704859,
    title = {{A state change in the low-mass X-ray binary XSS J12270−4859}},
    year = {2014},
    journal = {MNRAS},
    author = {Bassa, C G and Patruno, A and Hessels, J W T and Keane, E F and Monard, B and Mahony, E K and Bogdanov, S and Corbel, S and Edwards, P G and Archibald, A M and Janssen, G H and Stappers, B W and Tendulkar, S},
    number = {2},
    pages = {1825--1830},
    volume = {441},
    publisher = {Oxford University Press},
    url = {http://academic.oup.com/mnras/article/441/2/1825/1077796/A-state-change-in-the-lowmass-Xray-binary},
    doi = {10.1093/mnras/stu708},
    issn = {1365-2966}
}

@article{urpin1995,
    title = {{Accretion and evolution of the neutron star magnetic field}},
    year = {1995},
    journal = {MNRAS},
    author = {Urpin, V and Geppert, U},
    number = {4},
    pages = {1117--1124},
    volume = {275},
    url = {https://academic.oup.com/mnras/article-lookup/doi/10.1093/mnras/275.4.1117},
    doi = {10.1093/mnras/275.4.1117},
    issn = {0035-8711}
}

@article{Ertan2018,
    title = {{Accretion and propeller torque in the spin-down phase of neutron stars: The case of transitional millisecond pulsar PSR J1023+0038}},
    year = {2018},
    journal = {MNRAS},
    author = {Ertan, \"{U}nal},
    number = {1},
    month = {9},
    pages = {L12-L16},
    volume = {479},
    publisher = {Oxford Academic},
    url = {https://academic.oup.com/mnrasl/article/479/1/L12/4999905},
    doi = {10.1093/MNRASL/SLY089},
    issn = {1745-3925}
}

@article{Pringle1972AccretionSources,
    title = {{Accretion Disc Models for Compact X-Ray Sources}},
    year = {1972},
    journal = {Astronomy and Astrophysics, Vol. 21, p. 1 (1972)},
    author = {Pringle, J E and Rees, M J},
    pages = {1},
    volume = {21},
    url = {https://ui.adsabs.harvard.edu/abs/1972A%26A....21....1P/abstract},
    issn = {0004-6361}
}

@article{geppert1994,
    title = {{Accretion-driven magnetic field decay in neutron stars}},
    year = {1994},
    journal = {MNRAS},
    author = {Geppert, U and Urpin, V},
    number = {2},
    pages = {490--498},
    volume = {271},
    url = {https://academic.oup.com/mnras/article-lookup/doi/10.1093/mnras/271.2.490},
    isbn = {9788490225370},
    doi = {10.1093/mnras/271.2.490},
    issn = {0035-8711}
}

@article{Chen2017AnJ1808.43658,
    title = {{An evolutionary channel towards the accreting millisecond pulsar SAX J1808.4−3658}},
    year = {2017},
    journal = {Monthly Notices of the Royal Astronomical Society},
    author = {Chen, Wen Cong},
    number = {4},
    pages = {4673--4679},
    volume = {464},
    publisher = {Oxford Academic},
    url = {https://academic.oup.com/mnras/article/464/4/4673/2417467},
    doi = {10.1093/MNRAS/STW2747},
    issn = {0035-8711},
    arxivId = {1610.06767}
}

@article{Eggleton1983APPROXIMATIONSLOBES,
    title = {{APPROXIMATIONS TO THE RADII OF ROCHE LOBES}},
    year = {1983},
    journal = {The Astrophysical Journal},
    author = {Eggleton, Peter P},
    pages = {368--369},
    volume = {268}
}

@article{Shakura1973BlackAppearances,
    title = {{Black Holes in Binary Systems: Observational Appearances}},
    year = {1973},
    journal = {A{\&}A},
    author = {Shakura, N I and Sunyaev, R A},
    pages = {337},
    volume = {24},
    isbn = {9783540773405}
}

@incollection{Patruno2021AccretingPulsars,
    title = {{Accreting Millisecond X-Ray Pulsars}},
    year = {2021},
    booktitle = {Timing Neutron Stars: Pulsations, Oscillations and Explosions},
    author = {Patruno, A. and Watts, A. L.},
    editor = {Belloni Tomaso M.and M{\'{e}}ndez, Marianoand Zhang Chengmin},
    month = {6},
    pages = {143--208},
    publisher = {Springer Berlin Heidelberg},
    address = {Berlin, Heidelberg},
    isbn = {978-3-662-62110-3},
    issn = {2214-7985},
    arxivId = {1206.2727}
}

@article{Ghosh1979ACCRETION1,
    title = {{ACCRETION BY ROTATING MAGNETIC NEUTRON STARS. II. RADIAL AND VERTICAL STRUCTURE OF THE TRANSITION ZONE IN DISK ACCRETION 1}},
    year = {1979},
    journal = {ApJ},
    author = {Ghosh, P and Lamb, F K},
    pages = {259--276},
    volume = {232}
}

@article{Frank2002AccretionEdition,
    title = {{Accretion Power in Astrophysics: Third Edition}},
    year = {2002},
    journal = {Physics Today},
    author = {Frank, Juhan. and King, Andrew. and Raine, Derek.},
    pages = {398},
    volume = {39},
    url = {http://adsabs.harvard.edu/abs/2002apa..book.....F},
    isbn = {0521620538}
}

@inproceedings{DiSalvo2022AccretionPulsars,
    title = {{Accretion Powered X-ray Millisecond Pulsars}},
    year = {2022},
    booktitle = {Millisecond Pulsars},
    author = {Di Salvo, Tiziana and Sanna, Andrea},
    editor = {Bhattacharyya, Sudip and Papitto, Alessandro and Bhattacharya, Dipankar},
    month = {1},
    pages = {87--124},
    series = {Astrophysics and Space Science Library},
    volume = {465},
    publisher = {Springer-Verlag},
    address = {Berlin}
}

@article{Heinke2025CatalogBinaries,
    title = {{Catalog of Outbursts of Neutron Star Low-mass X-Ray Binaries}},
    year = {2025},
    journal = {ApJS},
    author = {Heinke, Craig O. and Zheng, Junwen and Maccarone, Thomas J. and Degenaar, Nathalie and Bahramian, Arash and Sivakoff, Gregory R. and Toor, Simrat},
    number = {2},
    month = {8},
    pages = {57},
    volume = {279},
    publisher = {IOP Publishing},
    url = {https://iopscience.iop.org/article/10.3847/1538-4365/ade99a https://iopscience.iop.org/article/10.3847/1538-4365/ade99a/meta},
    doi = {10.3847/1538-4365/ADE99A},
    issn = {0067-0049}
}

@article{Reimers1975CircumstellarGiants.,
    title = {{Circumstellar absorption lines and mass loss from red giants.}},
    year = {1975},
    journal = {MSRSL},
    author = {Reimers, D.},
    pages = {369--382},
    volume = {8},
    url = {https://ui.adsabs.harvard.edu/abs/1975MSRSL...8..369R/abstract}
}

@article{Van2021ConstrainingBraking,
    title = {{Constraining Progenitors of Observed Low-mass X-ray Binaries Using Convection and Rotation-Boosted Magnetic Braking}},
    year = {2021},
    journal = {ApJ},
    author = {Van, Kenny X. and Ivanova, Natalia},
    number = {2},
    month = {11},
    pages = {174},
    volume = {922},
    publisher = {IOP Publishing},
    url = {https://iopscience.iop.org/article/10.3847/1538-4357/ac236c https://iopscience.iop.org/article/10.3847/1538-4357/ac236c/meta},
    doi = {10.3847/1538-4357/AC236C},
    issn = {0004-637X}
}

@article{Halder2023DefiningPulsars,
    title = {{Defining Millisecond Pulsars}},
    year = {2023},
    journal = {Research Notes of the AAS},
    author = {Halder, Priyam and Goswami, Satyaki and Halder, Protyusha and Ghosh, Uday and Konar, Sushan},
    number = {10},
    month = {10},
    pages = {213},
    volume = {7},
    publisher = {IOP Publishing},
    url = {https://iopscience.iop.org/article/10.3847/2515-5172/ad00ac https://iopscience.iop.org/article/10.3847/2515-5172/ad00ac/meta},
    doi = {10.3847/2515-5172/AD00AC},
    issn = {2515-5172},
    arxivId = {2310.06230}
}

@article{Chevalier1991DiscoveryX-1.,
    title = {{Discovery of a 19-hour period in Aquila X-1.}},
    year = {1991},
    journal = {A{\&}A},
    author = {Chevalier, C. and Ilovaisky, S. A.},
    pages = {L11},
    volume = {251},
    url = {https://ui.adsabs.harvard.edu/abs/1991A&A...251L..11C/abstract},
    issn = {0004-6361}
}

@article{Markwardt2002DiscoveryJ1751305,
    title = {{Discovery of a Second Millisecond Accreting Pulsar: XTE J1751-305}},
    year = {2002},
    journal = {ApJL},
    author = {Markwardt, C.~B. and Swank, J.~H. and Strohmayer, T.~E. and in 't Zand, J.~J.~M. and Marshall, F.~E.},
    number = {1},
    pages = {L21-L24},
    volume = {575},
    doi = {10.1086/342612},
    arxivId = {astro-ph/astro-ph/0206491}
}

@article{Casella2008DiscoveryX-1,
    title = {{Discovery of Coherent Millisecond X-Ray Pulsations in Aquila X-1}},
    year = {2008},
    journal = {ApJ},
    author = {Casella, P and Altamirano, D and Patruno, A and Wijnands, R and van der Klis, M},
    number = {1},
    pages = {L41-L44},
    volume = {674},
    doi = {10.1086/528982},
    issn = {0004-637X},
    arxivId = {0708.1110}
}

@article{Bahramian2022,
    title = {{Low-Mass X-ray Binaries}},
    year = {2022},
    journal = {Phys. Scr.},
    author = {Bahramian, Arash and Degenaar, Nathalie},
    number = {T7},
    month = {6},
    pages = {87--93},
    volume = {1984},
    url = {https://arxiv.org/abs/2206.10053v1},
    doi = {10.48550/arxiv.2206.10053},
    issn = {14024896},
    arxivId = {2206.10053}
}

@article{konar1997,
    title = {{Magnetic field evolution of accreting neutron stars}},
    year = {1997},
    journal = {MNRAS},
    author = {Konar, Sushan and Bhattacharya, Dipankar},
    number = {2},
    pages = {311--317},
    volume = {284},
    url = {https://academic.oup.com/mnras/article-lookup/doi/10.1093/mnras/284.2.311},
    doi = {10.1093/mnras/284.2.311},
    issn = {0035-8711}
}

@article{Uzdensky2004MagneticDisks,
    title = {{Magnetic Interaction Between Stars And Accretion Disks}},
    year = {2004},
    journal = {Astrophysics and Space Science},
    author = {Uzdensky, Dmitri A},
    number = {1-4},
    pages = {573--585},
    volume = {292},
    url = {http://link.springer.com/10.1023/B:ASTR.0000045064.93078.87},
    doi = {10.1023/B:ASTR.0000045064.93078.87},
    issn = {0004-640X}
}

@article{Lovelace1999MagneticOutflows,
    title = {{Magnetic Propeller Outflows}},
    year = {1999},
    journal = {ApJ},
    author = {Lovelace, R V E and Romanova, M M and Bisnovatyi‐Kogan, G S},
    number = {1},
    pages = {368--372},
    volume = {514},
    url = {https://iopscience.iop.org/article/10.1086/306945},
    doi = {10.1086/306945},
    issn = {0004-637X}
}

@article{Radhakrishnan1982ONPULSAR,
    title = {{On the origin of the recently discovered ultra-rapid pulsar.}},
    year = {1982},
    journal = {Current Science},
    author = {Radhakrishnan, V and Srinivasan, G},
    number = {23},
    pages = {1096--1099},
    volume = {51},
    publisher = {Temporary Publisher},
    issn = {0011-3891}
}

@article{Ertan2021,
    title = {{On the torque reversals of accreting neutron stars}},
    year = {2021},
    journal = {MNRAS},
    author = {Ertan, \"{U}nal},
    number = {3},
    volume = {500},
    doi = {10.1093/mnras/staa3378},
    issn = {13652966}
}

@article{Kaaret2006DiscoveryJ1900.12455,
    title = {{Discovery of the Millisecond X‐Ray Pulsar HETE J1900.1−2455}},
    year = {2006},
    journal = {ApJ},
    author = {Kaaret, P. and Morgan, E. H. and Vanderspek, R. and Tomsick, J. A.},
    number = {2},
    month = {2},
    pages = {963--967},
    volume = {638},
    publisher = {American Astronomical Society},
    url = {https://ui.adsabs.harvard.edu/abs/2006ApJ...638..963K/abstract},
    doi = {10.1086/498886},
    issn = {0004-637X},
    arxivId = {astro-ph/0510483}
}

@article{Tauris2018DisentanglingLISA,
    title = {{Disentangling Coalescing Neutron Star-White Dwarf Binaries for LISA}},
    year = {2018},
    journal = {Physical Review Letters},
    author = {Tauris, Thomas M.},
    number = {13},
    month = {9},
    volume = {121},
    publisher = {American Physical Society},
    url = {http://arxiv.org/abs/1809.03504 http://dx.doi.org/10.1103/PhysRevLett.121.131105},
    doi = {10.1103/PhysRevLett.121.131105},
    arxivId = {1809.03504v2}
}

@article{Deng2021EvolutionPrescriptions,
    title = {{Evolution of LMXBs under Different Magnetic Braking Prescriptions}},
    year = {2021},
    journal = {ApJ},
    author = {Deng, Zhu-Ling and Li, Xiang-Dong and Gao, Zhi-Fu and Shao, Yong},
    pages = {174},
    volume = {909},
    url = {https://doi.org/10.3847/1538-4357/abe0b2},
    doi = {10.3847/1538-4357/abe0b2}
}

@article{Pylyser1988EvolutionComponent.,
    title = {{Evolution of low-mass close binary systems with a compact mass accreting component.}},
    year = {1988},
    journal = {A{\&}A},
    author = {Pylyser, E. H. P. and Savonije, G. J.},
    pages = {57--70},
    volume = {191},
    url = {https://ui.adsabs.harvard.edu/abs/1988A&A...191...57P/abstract},
    issn = {0004-6361}
}

@article{Podsiadlowski2002EvolutionaryBinaries,
    title = {{Evolutionary Sequences for Low‐ and Intermediate‐Mass X‐Ray Binaries}},
    year = {2002},
    journal = {ApJ},
    author = {Podsiadlowski, Ph. and Rappaport, S. and Pfahl, E. D.},
    number = {2},
    month = {2},
    pages = {1107--1133},
    volume = {565},
    publisher = {American Astronomical Society},
    url = {https://iopscience.iop.org/article/10.1086/324686 https://iopscience.iop.org/article/10.1086/324686/meta},
    doi = {10.1086/324686/FULLTEXT/},
    issn = {0004-637X}
}

@article{Van2019EvolvingBraking,
    title = {{Evolving LMXBs: CARB Magnetic Braking}},
    year = {2019},
    journal = {ApJL},
    author = {Van, Kenny X. and Ivanova, Natalia},
    number = {2},
    month = {11},
    pages = {L31},
    volume = {886},
    publisher = {IOP Publishing},
    url = {https://iopscience.iop.org/article/10.3847/2041-8213/ab571c https://iopscience.iop.org/article/10.3847/2041-8213/ab571c/meta},
    doi = {10.3847/2041-8213/AB571C},
    issn = {2041-8205},
    arxivId = {1911.05790}
}

@article{Bhattacharya1991FormationPulsars,
    title = {{Formation and evolution of binary and millisecond radio pulsars}},
    year = {1991},
    journal = {Physics Reports},
    author = {Bhattacharya, D. and van den Heuvel, E. P.J.},
    number = {1-2},
    month = {5},
    pages = {1--124},
    volume = {203},
    publisher = {North-Holland},
    doi = {10.1016/0370-1573(91)90064-S},
    issn = {0370-1573}
}

@article{He2019FormationPulsars,
    title = {{Formation of accreting millisecond X-ray pulsars}},
    year = {2019},
    journal = {Res. Astron. Astrophys.},
    author = {He, Xiang and Meng, Xiang-Cun and Chen, Hai-Liang},
    number = {8},
    month = {8},
    pages = {110},
    volume = {19},
    url = {https://iopscience.iop.org/article/10.1088/1674-4527/19/8/110},
    doi = {10.1088/1674-4527/19/8/110},
    issn = {1674-4527}
}

@article{Chen2013FormationPulsars.,
    title = {{Formation of black widows and redbacks—two distinct populations of eclipsing binary millisecond pulsars.}},
    year = {2013},
    journal = {ApJ},
    author = {Chen, Hai Liang and Chen, Xuefei and Tauris, Thomas M. and Han, Zhanwen},
    number = {1},
    month = {8},
    pages = {27},
    volume = {775},
    publisher = {IOP Publishing},
    url = {https://iopscience.iop.org/article/10.1088/0004-637X/775/1/27 https://iopscience.iop.org/article/10.1088/0004-637X/775/1/27/meta},
    doi = {10.1088/0004-637X/775/1/27},
    issn = {0004-637X},
    arxivId = {1308.4107}
}

@article{Heinke2013GalacticEvolution.,
    title = {{Galactic ultracompact x-ray binaries: disk stability and evolution.}},
    year = {2013},
    journal = {ApJ},
    author = {Heinke, C. O. and Ivanova, N. and Engel, M. C. and Pavlovskii, K. and Sivakoff, G. R. and Cartwright, T. F. and Gladstone, J. C.},
    number = {2},
    month = {4},
    pages = {184},
    volume = {768},
    publisher = {IOP Publishing},
    url = {https://iopscience.iop.org/article/10.1088/0004-637X/768/2/184 https://iopscience.iop.org/article/10.1088/0004-637X/768/2/184/meta},
    doi = {10.1088/0004-637X/768/2/184},
    issn = {0004-637X}
}

@article{Zdziarski2016IGRStar,
    title = {{IGR J17451-3022: Constraints on the nature of the donor star}},
    year = {2016},
    journal = {A{\&}A},
    author = {Zdziarski, Andrzej A. and Zi{\'{o}}{\l}kowski, Janusz and Bozzo, Enrico and Pjanka, Patryk},
    month = {11},
    pages = {A52},
    volume = {595},
    publisher = {EDP Sciences},
    url = {https://ui.adsabs.harvard.edu/abs/2016A&A...595A..52Z/abstract},
    doi = {10.1051/0004-6361/201628585},
    issn = {14320746},
    arxivId = {1603.07288}
}

@article{Misra2025InvestigatingEvolution,
    title = {{Investigating cannibalistic millisecond pulsar binaries using MESA: New constraints from pulsar spin and mass evolution}},
    year = {2025},
    journal = {A{\&}A},
    author = {Misra, Devina and Linares, Manuel and Ye, Claire S.},
    month = {1},
    pages = {A314},
    volume = {693},
    publisher = {EDP Sciences},
    url = {https://www.aanda.org/articles/aa/full_html/2025/01/aa52035-24/aa52035-24.html https://www.aanda.org/articles/aa/abs/2025/01/aa52035-24/aa52035-24.html},
    doi = {10.1051/0004-6361/202452035},
    issn = {0004-6361},
    arxivId = {2408.16048}
}

@article{Kar2024Long-termPulsars,
    title = {{Long-term evolution of spin and other properties of neutron star low-mass X-ray binaries: implications for millisecond X-ray pulsars}},
    year = {2024},
    journal = {MNRAS},
    author = {Kar, Abhijnan and Ojha, Pulkit and Bhattacharyya, Sudip},
    number = {1},
    month = {10},
    pages = {344--358},
    volume = {535},
    publisher = {Oxford Academic},
    url = {https://dx.doi.org/10.1093/mnras/stae2346},
    doi = {10.1093/MNRAS/STAE2346},
    issn = {0035-8711}
}

@article{Van2019Low-massPrescription,
    title = {{Low-mass X-ray binaries: the effects of the magnetic braking prescription}},
    year = {2019},
    journal = {MNRAS},
    author = {Van, K X and Ivanova, N and Heinke, C O},
    pages = {5595--5613},
    volume = {483},
    url = {https://academic.oup.com/mnras/article/483/4/5595/5256661},
    doi = {10.1093/mnras/sty3489}
}

@article{Gossage2023MagneticRegimes,
    title = {{Magnetic Braking with MESA Evolutionary Models in the Single Star and Low-mass X-Ray Binary Regimes}},
    year = {2023},
    journal = {ApJ},
    author = {Gossage, Seth and Kalogera, Vicky and Sun, Meng},
    number = {1},
    month = {6},
    pages = {27},
    volume = {950},
    publisher = {IOP Publishing},
    url = {https://iopscience.iop.org/article/10.3847/1538-4357/acc86e https://iopscience.iop.org/article/10.3847/1538-4357/acc86e/meta},
    doi = {10.3847/1538-4357/ACC86E},
    issn = {0004-637X}
}

@article{Konar1999MagneticIII,
    title = {{Magnetic field evolution of accreting neutron stars ± III}},
    year = {1999},
    journal = {MNRAS},
    author = {Konar, Sushan and Bhattacharya, Dipankar},
    pages = {795--798},
    volume = {308},
    url = {https://academic.oup.com/mnras/article/308/3/795/973644}
}

@article{Konar1999MagneticII,
    title = {{Magnetic {\textregistered}eld evolution of accreting neutron stars ± II}},
    year = {1999},
    journal = {MNRAS},
    author = {Konar, Sushan and Bhattacharya, Dipankar},
    pages = {588--594},
    volume = {303},
    url = {https://academic.oup.com/mnras/article/303/3/588/1002118},
    doi = {10.1046/j.1365-8711.1999.02287.x}
}

@article{Pavlovskii2016MassX-1,
    title = {{Mass transfer and magnetic braking in Sco X-1}},
    year = {2016},
    journal = {MNRAS},
    author = {Pavlovskii, K. and Ivanova, N.},
    number = {1},
    month = {2},
    pages = {263--269},
    volume = {456},
    publisher = {Oxford Academic},
    url = {https://dx.doi.org/10.1093/mnras/stv2685},
    doi = {10.1093/MNRAS/STV2685},
    issn = {0035-8711},
    arxivId = {1511.08847}
}

@article{Sanna2022MAXIJ1957+032:Binary,
    title = {{MAXI J1957+032: a new accreting millisecond X-ray pulsar in an ultra-compact binary}},
    year = {2022},
    journal = {MNRAS},
    author = {Sanna, A. and Bult, P. and Ng, M. and Ray, P. S. and Jaisawal, G. K. and Burderi, L. and Di Salvo, T. and Riggio, A. and Altamirano, D. and Strohmayer, T. E. and Manca, A. and Gendreau, K. C. and Chakrabarty, D. and Iwakiri, W. and Iaria, R.},
    number = {1},
    month = {9},
    pages = {L76-L80},
    volume = {516},
    publisher = {Oxford Academic},
    url = {https://dx.doi.org/10.1093/mnrasl/slac093},
    isbn = {59749.633146(18},
    doi = {10.1093/MNRASL/SLAC093},
    issn = {1745-3925},
    arxivId = {2208.05807}
}

@article{Paxton2011ModulesMESA,
    title = {{Modules for Experiments in Stellar Astrophysics (MESA)}},
    year = {2011},
    journal = {ApJS},
    author = {Paxton, Bill and Bildsten, Lars and Dotter, Aaron and Herwig, Falk and Lesaffre, Pierre and Timmes, Frank},
    number = {1},
    month = {1},
    pages = {3},
    volume = {192},
    url = {https://ui.adsabs.harvard.edu/abs/2011ApJS..192....3P/abstract},
    doi = {10.1088/0067-0049/192/1/3},
    issn = {00670049},
    arxivId = {1009.1622}
}

@article{Paxton2015ModulesExplosions,
    title = {{Modules for Experiments in Stellar Astrophysics (MESA): Binaries, pulsations, and explosions}},
    year = {2015},
    journal = {ApJS},
    author = {Paxton, Bill and Marchant, Pablo and Schwab, Josiah and Bauer, Evan B. and Bildsten, Lars and Cantiello, Matteo and Dessart, Luc and Farmer, R. and Hu, H. and Langer, N. and Townsend, R. H.D. and Townsley, Dean M. and Timmes, F. X.},
    number = {1},
    month = {9},
    pages = {15},
    volume = {220},
    publisher = {Institute of Physics Publishing},
    url = {https://ui.adsabs.harvard.edu/abs/2015ApJS..220...15P/abstract},
    doi = {10.1088/0067-0049/220/1/15},
    issn = {00670049},
    arxivId = {1506.03146}
}

@article{Paxton2018ModulesExplosions,
    title = {{Modules for Experiments in Stellar Astrophysics (MESA): Convective Boundaries, Element Diffusion, and Massive Star Explosions}},
    year = {2018},
    journal = {ApJS},
    author = {Paxton, Bill and Schwab, Josiah and Bauer, Evan B and Bildsten, Lars and Blinnikov, Sergei and Duffell, Paul and Farmer, R and Goldberg, Jared A and Marchant, Pablo and Sorokina, Elena and Thoul, Anne and Townsend, Richard H D and Timmes, F X},
    pages = {34},
    volume = {234},
    url = {https://doi.org/10.3847/1538-4365/aaa5a8},
    doi = {10.3847/1538-4365/aaa5a8}
}

@article{Paxton2013ModulesStars,
    title = {{Modules for experiments in stellar astrophysics (MESA): Planets, oscillations, rotation, and massive stars}},
    year = {2013},
    journal = {ApJS},
    author = {Paxton, Bill and Cantiello, Matteo and Arras, Phil and Bildsten, Lars and Brown, Edward F. and Dotter, Aaron and Mankovich, Christopher and Montgomery, M. H. and Stello, Dennis and Timmes, F. X. and Townsend, Richard},
    number = {1},
    month = {9},
    pages = {4},
    volume = {208},
    url = {https://ui.adsabs.harvard.edu/abs/2013ApJS..208....4P/abstract},
    doi = {10.1088/0067-0049/208/1/4},
    issn = {00670049},
    arxivId = {1301.0319}
}

@article{Paxton2019ModulesConservation,
    title = {{Modules for Experiments in Stellar Astrophysics (MESA): Pulsating Variable Stars, Rotation, Convective Boundaries, and Energy Conservation}},
    year = {2019},
    journal = {ApJS},
    author = {Paxton, Bill and Smolec, R and Schwab, Josiah and Gautschy, A and Bildsten, Lars and Cantiello, Matteo and Dotter, Aaron and Farmer, R and Goldberg, Jared A and Jermyn, Adam S and Kanbur, S M and Marchant, Pablo and Thoul, Anne and Townsend, Richard H D and Wolf, William M and Zhang, Michael and Timmes, F X},
    pages = {10},
    volume = {243},
    url = {https://doi.org/10.3847/1538-4365/ab2241},
    doi = {10.3847/1538-4365/ab2241}
}

@article{Jermyn2023ModulesInfrastructure,
    title = {{Modules for Experiments in Stellar Astrophysics (MESA): Time-dependent Convection, Energy Conservation, Automatic Differentiation, and Infrastructure}},
    year = {2023},
    journal = {ApJS},
    author = {Jermyn, Adam S and Bauer, Evan B and Schwab, Josiah and Farmer, R and Ball, Warrick H and Bellinger, Earl P and Dotter, Aaron and Joyce, Meridith and Marchant, Pablo and Mombarg, Joey S G and Bildsten, Lars and Townsend, Richard H D and Timmes, F X},
    pages = {15},
    volume = {265},
    url = {https://doi.org/10.3847/1538-4365/acae8d},
    doi = {10.3847/1538-4365/acae8d}
}

@article{Sengar2017NovelStars,
    title = {{Novel modelling of ultracompact X-ray binary evolution-stable mass transfer from white dwarfs to neutron stars}},
    year = {2017},
    journal = {MNRAS},
    author = {Sengar, Rahul and Tauris, Thomas M and Langer, Norbert and Istrate, Alina G},
    pages = {6--10},
    volume = {470},
    url = {https://academic.oup.com/mnrasl/article/470/1/L6/3778283},
    doi = {10.1093/mnrasl/slx064}
}

@article{Bildsten1997ObservationsPulsars,
    title = {{Observations of Accreting Pulsars}},
    year = {1997},
    journal = {ApJS},
    author = {Bildsten, Lars and Chakrabarty, Deepto and Chiu, John and Finger, Mark H. and Koh, Danny T. and Nelson, Robert W. and Prince, Thomas A. and Rubin, Bradley C. and Scott, D. Matthew and Stollberg, Mark and Vaughan, Brian A. and Wilson, Colleen A. and Wilson, Robert B.},
    number = {2},
    month = {12},
    pages = {367},
    volume = {113},
    publisher = {IOP Publishing},
    url = {https://iopscience.iop.org/article/10.1086/313060 https://iopscience.iop.org/article/10.1086/313060/meta},
    doi = {10.1086/313060},
    issn = {0067-0049},
    arxivId = {astro-ph/9707125}
}

@article{Niang2024OnBinaries,
    title = {{On the lack of X-ray pulsation in most neutron star low-mass X-ray binaries}},
    year = {2024},
    journal = {MNRAS},
    author = {Niang, N. and Ertan, \"{U} and Gen{\c{c}}ali, A. A. and Toyran, O. and Ulubay, A. and Devlen, E. and Alpar, M. A. and G{\"{u}}gercinoǧlu, E.},
    number = {2},
    month = {7},
    pages = {2133--2142},
    volume = {532},
    publisher = {Oxford Academic},
    url = {https://dx.doi.org/10.1093/mnras/stae1595},
    doi = {10.1093/MNRAS/STAE1595},
    issn = {0035-8711},
    arxivId = {2406.17921}
}

@article{Urpin1998OnPulsars,
    title = {{On the origin of millisecond pulsars}},
    year = {1998},
    journal = {A{\&}A},
    author = {Urpin, V. and Geppert, U. and Konenkov, D.},
    month = {3},
    pages = {244},
    volume = {331},
    url = {https://ui.adsabs.harvard.edu/abs/1998A%26A...331..244U/abstract},
    issn = {0004-6361}
}

@article{Srinivasan2010,
    title = {{Recycled pulsars}},
    year = {2010},
    journal = {New Astron. Rev.},
    author = {Srinivasan, G.},
    number = {3-6},
    month = {3},
    pages = {93--100},
    volume = {54},
    publisher = {North-Holland},
    doi = {10.1016/j.newar.2010.09.026},
    issn = {13876473}
}

@article{Tauris2012,
    title = {{Spin-down of radio millisecond pulsars at genesis.}},
    year = {2012},
    journal = {Science},
    author = {Tauris, Thomas M},
    number = {6068},
    month = {2},
    pages = {561--3},
    volume = {335},
    publisher = {American Association for the Advancement of Science},
    url = {http://www.ncbi.nlm.nih.gov/pubmed/22301314},
    doi = {10.1126/science.1216355},
    issn = {1095-9203},
    pmid = {22301314}
}

@article{Galloway2020,
    title = {{The Multi-INstrument Burst ARchive (MINBAR)}},
    year = {2020},
    journal = {ApJS},
    author = {Galloway, Duncan K. and in ’t Zand, Jean and Chenevez, Jérôme and W{\"{o}}rpel, Hauke and Keek, Laurens and Ootes, Laura and Watts, Anna L. and Gisler, Luis and Sanchez-Fernandez, Celia and Kuulkers, Erik},
    number = {2},
    month = {8},
    pages = {32},
    volume = {249},
    url = {https://iopscience.iop.org/article/10.3847/1538-4365/ab9f2e},
    doi = {10.3847/1538-4365/ab9f2e},
    issn = {1538-4365}
}

@article{Gencali2022,
    title = {{The torque reversals of 4U 1626–67}},
    year = {2022},
    journal = {A{\&}A},
    author = {Gen{\c{c}}ali, A. A. and Niang, N and Toyran, O. and Ertan, \"{U}. and Ulubay, A. and {\c{S}}a{\c{s}}maz, S. and Devlen, E. and Vahdat, A. and {\"{O}}zcan, Ş. and Alpar, M. A.},
    month = {2},
    pages = {A13},
    volume = {658},
    publisher = {EDP Sciences},
    url = {https://www.aanda.org/articles/aa/full_html/2022/02/aa41772-21/aa41772-21.html https://www.aanda.org/articles/aa/abs/2022/02/aa41772-21/aa41772-21.html},
    doi = {10.1051/0004-6361/202141772},
    issn = {0004-6361},
    arxivId = {2110.15392}
}

@book{Tauris2023PhysicsSources,
    title = {{Physics of Binary Star Evolution. From Stars to X-ray Binaries and Gravitational Wave Sources}},
    year = {2023},
    author = {Tauris, Thomas M and van den Heuvel, Edward P.~J.},
    pages = {864},
    publisher = {Princeton University Press}
}

@article{Vanderspek2005PossibleHETE,
    title = {{Possible new X-ray burst source detected by HETE}},
    year = {2005},
    journal = {ATel},
    author = {Vanderspek, Roland and Morgan, Ed and Crew, Geoff and Graziani, Carlo and Suzuki, Motoko and Vanderspek, Roland and Morgan, Ed and Crew, Geoff and Graziani, Carlo and Suzuki, Motoko},
    pages = {1},
    volume = {516},
    url = {https://ui.adsabs.harvard.edu/abs/2005ATel..516....1V/abstract}
}

@article{Riggio2011SecularJ1751-305,
    title = {{Secular spin-down of the AMP XTE J1751-305}},
    year = {2011},
    journal = {A{\&}A},
    author = {Riggio, A. and Burderi, L. and Salvo, T. Di and Papitto, A. and D’A{\`{i}}, A. and Iaria, R. and Menna, M. T.},
    month = {7},
    pages = {A140},
    volume = {531},
    publisher = {EDP Sciences},
    url = {https://www.aanda.org/articles/aa/full_html/2011/07/aa14883-10/aa14883-10.html https://www.aanda.org/articles/aa/abs/2011/07/aa14883-10/aa14883-10.html},
    doi = {10.1051/0004-6361/201014883},
    issn = {0004-6361}
}

@article{Papitto2013SwingsPulsar,
    title = {{Swings between rotation and accretion power in a binary millisecond pulsar}},
    year = {2013},
    journal = {Nature},
    author = {Papitto, A. and Ferrigno, C. and Bozzo, E. and Rea, N. and Pavan, L. and Burderi, L. and Burgay, M. and Campana, S. and Di Salvo, T. and Falanga, M. and Filipovi{\'{c}}, M. D. and Freire, P. C.C. and Hessels, J. W.T. and Possenti, A. and Ransom, S. M. and Riggio, A. and Romano, P. and Sarkissian, J. M. and Stairs, I. H. and Stella, L. and Torres, D. F. and Wieringa, M. H. and Wong, G. F.},
    number = {7468},
    month = {9},
    pages = {517--520},
    volume = {501},
    publisher = {Nature Publishing Group},
    url = {https://www.nature.com/articles/nature12470},
    doi = {10.1038/nature12470},
    issn = {1476-4687}
}

@article{Deloye2008ThePerspective,
    title = {{The Connection Between Low‐Mass X‐ray Binaries and (Millisecond) Pulsars: A Binary Evolution Perspective}},
    year = {2008},
    journal = {AIP Conference Proceedings},
    author = {Deloye, Christopher J.},
    number = {1},
    month = {2},
    pages = {501--509},
    volume = {983},
    publisher = {AIP Publishing},
    url = {/aip/acp/article/983/1/501/620989/The-Connection-Between-Low-Mass-X-ray-Binaries-and},
    isbn = {9780735405028},
    doi = {10.1063/1.2900285},
    issn = {0094-243X}
}

@article{Lasota2001TheTransients,
    title = {{The disc instability model of dwarf novae and low-mass X-ray binary transients}},
    year = {2001},
    journal = {New Astronomy Reviews},
    author = {Lasota, Jean Pierre},
    number = {7},
    pages = {449--508},
    volume = {45},
    publisher = {Elsevier},
    url = {https://ui.adsabs.harvard.edu/abs/2001NewAR..45..449L/abstract},
    doi = {10.1016/S1387-6473(01)00112-9},
    issn = {13876473},
    arxivId = {astro-ph/0102072}
}

@article{Sanchez2017TheSpectroscopy,
    title = {{The donor of Aquila X-1 revealed by high-angular resolution near-infrared spectroscopy}},
    year = {2017},
    journal = {MNRAS},
    author = {S{\'{a}}nchez, D. Mata and Mu{\~{n}}oz-Darias, T. and Casares, J. and Jim{\'{e}}nez-Ibarra, F.},
    number = {1},
    month = {1},
    pages = {L41-L45},
    volume = {464},
    publisher = {Oxford Academic},
    url = {https://dx.doi.org/10.1093/mnrasl/slw172},
    doi = {10.1093/MNRASL/SLW172},
    issn = {1745-3925},
    arxivId = {1609.00392}
}

@article{Lan2023ThePulsars,
    title = {{The Effect of Irradiation on the Spin of Millisecond Pulsars}},
    year = {2023},
    journal = {ApJL},
    author = {Lan, Shun-Yi and Meng, Xiang-Cun},
    number = {1},
    month = {10},
    pages = {L24},
    volume = {956},
    publisher = {IOP Publishing},
    url = {https://iopscience.iop.org/article/10.3847/2041-8213/acfedf https://iopscience.iop.org/article/10.3847/2041-8213/acfedf/meta},
    doi = {10.3847/2041-8213/ACFEDF},
    issn = {2041-8205}
}

@article{Bhattacharyya2017THEPULSARS,
    title = {{THE EFFECT OF TRANSIENT ACCRETION ON THE SPIN-UP OF MILLISECOND PULSARS}},
    year = {2017},
    journal = {The Astrophysical Journal},
    author = {Bhattacharyya, Sudip and Chakrabarty, Deepto},
    number = {1},
    month = {1},
    pages = {4},
    volume = {835},
    publisher = {IOP Publishing},
    url = {https://iopscience.iop.org/article/10.3847/1538-4357/835/1/4 https://iopscience.iop.org/article/10.3847/1538-4357/835/1/4/meta},
    doi = {10.3847/1538-4357/835/1/4},
    issn = {0004-637X}
}

@article{Pylyser1989TheLosses.,
    title = {{The evolution of low-mass close binary systems with a compact component. II. Systems captured by angular momentum losses.}},
    year = {1989},
    journal = {A{\&}A},
    author = {Pylyser, E. H. P. and Savonije, G. J.},
    pages = {52--62},
    volume = {208},
    url = {https://ui.adsabs.harvard.edu/abs/1989A&A...208...52P/abstract},
    issn = {0004-6361}
}

@article{Istrate2014ThePeriod,
    title = {{The formation of low-mass helium white dwarfs orbiting pulsars - Evolution of low-mass X-ray binaries below the bifurcation period}},
    year = {2014},
    journal = {A{\&}A},
    author = {Istrate, A. G. and M Tauris, T. and Langer, N.},
    month = {11},
    pages = {A45},
    volume = {571},
    publisher = {EDP Sciences},
    url = {https://www.aanda.org/articles/aa/full_html/2014/11/aa24680-14/aa24680-14.html https://www.aanda.org/articles/aa/abs/2014/11/aa24680-14/aa24680-14.html},
    doi = {10.1051/0004-6361/201424680},
    issn = {0004-6361},
    arxivId = {1410.5470}
}

@article{Echeveste2024TheLMXBs,
    title = {{The impact of different magnetic braking prescriptions on the evolution of LMXBs}},
    year = {2024},
    journal = {MNRAS},
    author = {Echeveste, M. and Novarino, M. L. and Benvenuto, O. G. and De Vito, M. A.},
    month = {4},
    url = {https://ui.adsabs.harvard.edu/abs/2024MNRAS.tmp.1119E/abstract},
    doi = {10.1093/MNRAS/STAE1115},
    issn = {0035-8711},
    arxivId = {arXiv:2404.16185}
}

@article{Yang2024TheBinaries,
    title = {{The influence of the magnetic braking laws on the evolution of persistent and transient low-mass X-ray binaries}},
    year = {2024},
    journal = {ApJ},
    author = {Yang, Hao-Ran and Li, Xiang-Dong},
    number = {2},
    month = {10},
    pages = {298},
    volume = {974},
    url = {https://iopscience.iop.org/article/10.3847/1538-4357/ad7824},
    doi = {10.3847/1538-4357/AD7824},
    issn = {0004-637X},
    arxivId = {2409.05067}
}

@article{Ertan2017TheStars,
    title = {{The inner disc radius in the propeller phase and accretion-propeller transition of neutron stars}},
    year = {2017},
    journal = {MNRAS},
    author = {Ertan, \"{U}nal},
    number = {1},
    month = {4},
    pages = {175--180},
    volume = {466},
    publisher = {Oxford Academic},
    url = {https://academic.oup.com/mnras/article/466/1/175/2638369},
    doi = {10.1093/MNRAS/STW3131},
    issn = {0035-8711}
}

@article{Misra2025TheTransfer,
    title = {{The slowest spinning Galactic-field spider PSR J1932+2121: a history of inefficient mass transfer}},
    year = {2025},
    journal = {MNRAS},
    author = {Misra, Devina and Koljonen, Karri I.I. and Linares, Manuel},
    number = {1},
    month = {5},
    pages = {L58-L64},
    volume = {541},
    publisher = {Oxford Academic},
    url = {https://dx.doi.org/10.1093/mnrasl/slaf054},
    doi = {10.1093/MNRASL/SLAF054},
    issn = {1745-3925},
    arxivId = {2504.05372}
}

@article{Galloway2008ThermonuclearExplorer,
    title = {{Thermonuclear (Type I) X‐Ray Bursts Observed by the Rossi X‐Ray Timing Explorer}},
    year = {2008},
    journal = {ApJS},
    author = {Galloway, Duncan K. and Muno, Michael P. and Hartman, Jacob M. and Psaltis, Dimitrios and Chakrabarty, Deepto},
    number = {2},
    month = {12},
    pages = {360--422},
    volume = {179},
    url = {https://iopscience.iop.org/article/10.1086/592044},
    doi = {10.1086/592044},
    issn = {0067-0049}
}

@article{Skumanich1972TimeDepletion,
    title = {{Time Scales for Ca II Emission Decay, Rotational Braking, and Lithium Depletion}},
    year = {1972},
    journal = {ApJ},
    author = {Skumanich, A.},
    month = {2},
    pages = {565},
    volume = {171},
    publisher = {American Astronomical Society},
    url = {https://ui.adsabs.harvard.edu/abs/1972ApJ...171..565S/abstract},
    doi = {10.1086/151310},
    issn = {0004-637X}
}

@incollection{Papitto2022TransitionalPulsars,
    title = {{Transitional Millisecond Pulsars}},
    year = {2022},
    booktitle = {Millisecond Pulsars},
    author = {Papitto, Alessandro and de Martino, Domitilla},
    editor = {Bhattacharyya, Sudip and Papitto, Alessandro and Bhattacharya, Dipankar},
    pages = {157--200},
    publisher = {Springer International Publishing},
    address = {Cham}
}

@article{Fromang2009TurbulentInstability,
    title = {{Turbulent resistivity driven by the magnetorotational instability}},
    year = {2009},
    journal = {Astronomy {\&} Astrophysics},
    author = {Fromang, S. and Stone, J. M.},
    number = {1},
    month = {11},
    pages = {19--28},
    volume = {507},
    publisher = {EDP Sciences},
    url = {https://www.aanda.org/articles/aa/abs/2009/43/aa12752-09/aa12752-09.html},
    doi = {10.1051/0004-6361/200912752},
    issn = {0004-6361},
    arxivId = {0906.4422}
}

@article{Faulkner1971Ultrashort-PeriodCamelopardalis,
    title = {{Ultrashort-Period Binaries, Gravitational Radiation, and Mass Transfer. I. The Standard Model, with Applications to WZ Sagittae and Z Camelopardalis}},
    year = {1971},
    journal = {ApJL},
    author = {Faulkner, John},
    month = {12},
    pages = {L99},
    volume = {170},
    publisher = {American Astronomical Society},
    url = {https://ui.adsabs.harvard.edu/abs/1971ApJ...170L..99F/abstract},
    doi = {10.1086/180848},
    issn = {0004-637X}
}

@article{Dubus1999,
    title = {{X-ray irradiation in low-mass binary systems}},
    year = {1999},
    journal = {MNRAS},
    author = {Dubus, Guillaume and Lasota, Jean Pierre and Hameury, Jean Marie and Charles, Phil},
    number = {1},
    month = {2},
    pages = {139--147},
    volume = {303},
    publisher = {Blackwell Publishing Ltd},
    url = {https://academic.oup.com/mnras/article/303/1/139/1140300},
    doi = {10.1046/J.1365-8711.1999.02212.X/2/303-1-139-FIG010.JPEG},
    issn = {00358711},
    arxivId = {astro-ph/9809036}
}

@article{Deloye2003WhiteObjects,
    title = {{White Dwarf Donors in Ultracompact Binaries: The Stellar Structure of Finite‐Entropy Objects}},
    year = {2003},
    journal = {ApJ},
    author = {Deloye, Christopher J. and Bildsten, Lars},
    number = {2},
    month = {12},
    pages = {1217--1228},
    volume = {598},
    publisher = {American Astronomical Society},
    url = {https://iopscience.iop.org/article/10.1086/379063 https://iopscience.iop.org/article/10.1086/379063/meta},
    doi = {10.1086/379063/FULLTEXT/},
    issn = {0004-637X}
}

@article{Avakyan2023XRBcats:Catalogue,
    title = {{XRBcats: Galactic low-mass X-ray binary catalogue}},
    year = {2023},
    journal = {A{\&}A},
    author = {Avakyan, A. and Neumann, M. and Zainab, A. and Doroshenko, V. and Wilms, J. and Santangelo, A.},
    month = {7},
    pages = {A199},
    volume = {675},
    publisher = {EDP Sciences},
    url = {https://www.aanda.org/articles/aa/full_html/2023/07/aa46522-23/aa46522-23.html https://www.aanda.org/articles/aa/abs/2023/07/aa46522-23/aa46522-23.html},
    doi = {10.1051/0004-6361/202346522},
    issn = {0004-6361},
    arxivId = {2303.16168}
}

@article{Ustyugova2006PropellerStars,
    title = {{“Propeller” Regime of Disk Accretion to Rapidly Rotating Stars}},
    year = {2006},
    journal = {ApJ},
    author = {Ustyugova, G. V. and Koldoba, A. V. and Romanova, M. M. and Lovelace, R. V. E.},
    number = {1},
    month = {7},
    pages = {304--318},
    volume = {646},
    publisher = {American Astronomical Society},
    url = {https://iopscience.iop.org/article/10.1086/503379 https://iopscience.iop.org/article/10.1086/503379/meta},
    doi = {10.1086/503379/FULLTEXT/},
    issn = {0004-637X},
    arxivId = {astro-ph/0603249}
}





\bsp	
\label{lastpage}
\end{document}